\documentclass[useAMS]{mn2e}
\usepackage{times}
\usepackage{txfonts}
\usepackage{bm}

\usepackage{graphicx}
\usepackage[dvips]{color}

\newcommand\plotone[1]{\centering\includegraphics[width=\hsize]{#1}}

\newcommand\gessim{\mathrel {\vcenter {\baselineskip 0pt \kern 0pt \hbox{$>$} \kern 0pt \hbox{$\sim$} }}}
\newcommand\kms{{\rm \,km\,s^{-1}}}
\newcommand\beq{\begin{equation}}
\newcommand\eeq{\end{equation}}
\newcommand\ve{v_{\rm esc}}
\newcommand\vc{v_{\rm circ}}
\newcommand\vv{v_{\rm vir}}
\newcommand\vmin{v_{\rm min}}
\newcommand\vr{v_{\rm r}}
\newcommand\rapo{r_{\rm apo}}
\newcommand\sm{{\rm M}_\odot}
\title{The RAVE Survey: Constraining the Local Galactic Escape Speed}

\author[Smith et al.]{
Martin C. Smith,$^1$\thanks{msmith@astro.rug.nl}
Gregory R.~Ruchti,$^2$
Amina Helmi,$^1$
Rosemary F.G. Wyse,$^2$
\newauthor
J.P. Fulbright,$^2$
K.C. Freeman,$^3$
J.F. Navarro,$^4$
G.M. Seabroke,$^5$
M. Steinmetz,$^6$
\newauthor
M. Williams,$^3$
O. Bienaym\'e,$^7$
J. Binney,$^8$
J. Bland-Hawthorn,$^9$
W. Dehnen,$^{10}$
\newauthor
B. K. Gibson,$^{11}$
G. Gilmore,$^5$
E.K. Grebel,$^{12}$
U. Munari,$^{13}$
Q.A. Parker,$^{9,14}$
\newauthor
R.-D. Scholz,$^6$
A. Siebert,$^6$
F.G. Watson,$^9$
T. Zwitter$^{15}$
\\\\
$^1$Kapteyn Astronomical Institute, University of Groningen, P.O. Box 800, 9700 AV Groningen, The Netherlands\\
$^2$Johns Hopkins University, 366 Bloomberg Center, 3400 North Charles 
Street, Baltimore, MD 21218, USA\\
$^3$Research School of Astronomy and Astrophysics, Mount Stromlo Observatory, Cotter Road, Weston Creek, Canberra, ACT 72611, Australia.\\
$^4$University of Victoria, P.O. Box 3055, Station CSC, Victoria, BC V8W 3P6, Canada\\
$^5$Institute of Astronomy, University of Cambridge, Madingley Road, Cambridge CB3 0HA, UK\\
$^6$Astrophysikalisches Institut Potsdam, An der Sterwarte 16, D-14482 
Potsdam, Germany\\
$^7$Observatoire de Strasbourg, 11 Rue de L'Universit\'{e}, 67000 Strasbourg, 
France\\
$^8$Rudolf Peierls Centre for Theoretical Physics, University of Oxford, 1 
Keble Road, Oxford OX1 3NP, UK\\
$^9$Anglo-Australian Observatory, P.O. Box 296, Epping, NSW 1710, Australia\\
$^{10}$University of Leicester, University Road, Leicester LE1 7RH, UK\\
$^{11}$University of Central Lancashire, Preston PR1 2HE, UK\\
$^{12}$Astronomisches Institut, Universit\"{a}t Basel, Venusstrasse 7,
Binningen CH-4102, Switzerland\\
$^{13}$INAF Osservatorio Astronomico di Padova, Via dell'Osservatorio 8, Asiago I-36012, Italy\\
$^{14}$Macquarie University, Sydney, NSW 2109, Australia\\
$^{15}$Department of Physics, University of Ljubljana, Jadranska 19, Ljubljana, Slovenia
}

\date{
Accepted ........
Received .......;
in original form ......}

\pubyear{2005}

\begin{document}
\maketitle

\begin{abstract}
We report new constraints on the local escape speed of our Galaxy. Our
analysis is based on a sample of high velocity stars from the RAVE
survey and two previously published datasets. We use cosmological
simulations of disk galaxy formation to motivate our assumptions on
the shape of the velocity distribution, allowing for a significantly
more precise measurement of the escape velocity compared to previous
studies. We find that the escape velocity lies within the range $498\kms
< \ve < 608 \kms$ (90 per cent confidence), with a median likelihood
of $544\kms$.  The fact that $\ve^2$ is significantly greater than
$2\vc^2$ (where $\vc=220\kms$ is the local circular velocity) implies
that there must be a significant amount of mass exterior to the Solar
circle, i.e. this convincingly demonstrates the presence of a dark
halo in the Galaxy. 
We use our constraints on $\ve$ to determine the mass of the Milky Way
halo for three halo profiles. For example, an adiabatically contracted NFW
halo model results in a virial mass of
$1.42^{+1.14}_{-0.54}\times10^{12}M_\odot$ and virial radius of 
$305^{+66}_{-45}$ kpc (90 per cent confidence).
For this model the circular velocity at the virial radius is
$142^{+31}_{-21}\kms$.
Although our halo masses
are model dependent, we find that they are in good agreement with each other.
\end{abstract}


\section{Introduction}
\label{sec:intro}

The existence of a dark halo around the Milky Way has been known for
many years, although its nature is still uncertain. The mass and
extent of this halo is a significant issue in astronomy. One vital
tool which can be used to tackle this problem is
also one of the simplest - the escape speed. If we are able to
determine the escape speed at the solar neighbourhood, i.e. the
velocity that a star requires to escape the local gravitational
field of the Milky Way, then this can provide important constraints on
the extent of the dark halo. The reason why this quantity is so
important is because it is the only local dynamical measurement that
can be used to probe the extent of the mass distribution outside the
solar circle.
Unlike the circular velocity, which depends primarily on the mass
interior to the solar circle, the escape velocity contains information
about the mass exterior to the solar circle. Although one needs a
model for this mass distribution, the escape velocity (i.e. the local
gravitational potential) can be used as a constraint from which it is
possible to determine the total mass.

It is possible to use more distant measurements to investigate the
extent of the Galactic halo.
Unfortunately, gas rotation curves cannot be traced beyond the extent
of gas in circular orbits, $\sim20$ kpc for the Milky Way. The task of
tracing the rotation curve is also complicated by the fact that
velocities have to be accompanied by distances (Binney \& Dehnen 1997)
and, in any case, our Galaxy does not appear to have an extended HI disk.
As a consequence, most methods of probing the halo rely on satellites
and globular clusters, whose velocities can be measured out to
significantly greater Galactocentric distances. Many authors have used
the velocities of the Milky Way's satellite galaxies and globular
clusters in an attempt 
to constrain the total halo mass. Although numerous papers have
dealt with this subject (Little \& Tremaine 1987; Zaritsky et
al. 1989; Kulessa \& Lynden-Bell 1992; Kochanek 1996), two of the more
recent ones to exploit the motions of satellite galaxies and
globular clusters have concluded the total mass of the halo to be
around $2~\times 10^{12}\, {\rm M}_\odot$ : Wilkinson \& Evans
(1999) found a halo mass of $\sim 1.9^{+3.6}_{-1.7} \times
10^{12}~{\rm M}_\odot$ by adopting a halo model which produces a flat
rotation curve that is truncated beyond an outer edge; whereas
Sakamoto, Chiba \& Beers (2003), using a halo potential that gives a
flat rotation curve, also included the velocities of field
horizontal-branch stars to find a total halo mass of
$2.5^{+0.5}_{-1.0} \times 10^{12}~{\rm M}_\odot$ or $1.8^{+0.4}_{-0.7}
\times 10^{12}~{\rm M}_\odot$, depending on whether or not the
analysis includes Leo I. Another complementary approach that can be
adopted is to analyse the radial velocity dispersion profile of halo
objects; Battaglia et al. (2005; 2006) used this method to determine a total
mass of $0.5-1.5 \times 10^{12}~{\rm M}_\odot$ depending on the chosen
model for the halo profile (see also Dehnen, McLaughlin \& Sachania
(2006) for a reanalysis of this dataset).
The future for this field looks
promising with space missions such as Gaia (due for launch 2011; Perryman et
al. 2001; Wilkinson et al. 2006) and Space Interferometery Mission
(due for launch $\sim2015$; Allen, Shao \& Peterson 1998), since such
missions will be able to provide accurate proper motion information to
complement the existing radial velocity measurements; with such high
quality data it should be possible to determine the mass of the Milky
Way to $\sim20$ per cent (Wilkinson \& Evans 1999).

One can see that the current results mentioned above still produce a
factor of $\sim2$ uncertainty in the mass of the Milky Way, due to the
fact that the results are still model dependent and are hindered by
small number statistics concerning the relevant datasets. Therefore it
would be very valuable if one could provide tight constraints on the
local escape velocity in order to pin down the gravitational potential
at this point. As far back as the 1920s samples of high velocity stars
were being used to estimate the local escape velocity (for example,
Oort 1926; Oort 1928). As the 20th century progressed, many papers
refined the estimate of $\ve$ (see Fich \& Tremaine [1991] for a
review), culminating in the final decade with the seminal work of
Leonard \& Tremaine (1990, hereafter LT90) and the subsequent 
refinement by  Kochanek (1996, hereafter K96). These two papers
concluded that, to 90 per cent confidence, the escape velocity lies in
the range $450\kms < \ve < 650 \kms$ and $489\kms < \ve < 730 \kms$,
respectively. Their conclusions are hampered by several problems:
firstly, the paucity of high velocity stars from which to estimate
$\ve$; secondly the fact that biases were introduced by selecting high
velocity stars from proper-motion surveys; and thirdly the uncertainty
in the assumptions regarding the underlying form of the tail of the
velocity distribution. In this new century the difficulties posed by
the first two issues are to some extent diminishing due to the large
kinematically unbiased surveys that are now underway or planned,
such as RAdial Velocity Experiment (RAVE; Steinmetz et al. 2006; see
also Section \ref{sec:rave}), Sloan Extension for Galactic
Understanding and Exploration (SEGUE; Beers et al. 2004) and Gaia
(Perryman et al. 2001). The latter problem can be tackled through
various methods; one such approach could be to use predictions from
cosmological simulations to estimate the form of the velocity
distributions. In this paper we shall make use of the advancement
afforded to us by the RAVE survey, combined with the analysis of
cosmological simulations, to refine the determination of $\ve$.

The outline of this paper is as follows. In Section \ref{sec:analysis}
we review the analytical techniques that have been developed to
constrain the escape velocity from a sample of velocity
measurements. Then in the following section we assess various aspects
of these techniques using cosmological simulations. In particular we
use the simulations to estimate the expected shape of the tail of the
velocity distribution, which is a crucial ingredient in the escape
velocity analysis. In Section \ref{sec:data} we present the data that
we will use to constrain the escape velocity and undertake some tests
to ensure that these data are reliable. Our new data come from the
RAVE project (Steinmetz et al. 2006), but are augmented with
archival data from published surveys. In Section
\ref{sec:results} we present our results and in Section \ref{sec:disc}
we consider some of the issues arising from or concerning these
results; in particular, this latter section discusses the nature of
our high velocity stars (Section \ref{sec:rpmd}), the effect of the
sample volume on the recovery of the escape velocity
(\ref{sec:volume}), the possible contamination from unbound stars
(\ref{sec:unbound}) and also uses our new constraints on $\ve$ to
investigate the total mass of the Galactic halo
(\ref{sec:halomass}). In Section \ref{sec:conclusion} we conclude our
paper with a brief summary.

\section{Analysis techniques}
\label{sec:analysis}

\subsection{Likelihood}
\label{sec:formal}

The techniques that we apply in order to constrain the escape velocity
($\ve$) are based on those established by LT90. They parametrize the
distribution of stellar velocities around $\ve$ according to the
following formula,
\beq
\label{eq:ltdist}
f(|{\bf v}|\,|\,\ve,k) \propto (\ve-|{\bf v}|)^{k}, ~~~ |{\bf v}| <
\ve \eeq \beq f(|{\bf v}|\,|\,\ve,k) =0, ~~~ |{\bf v}| \ge \ve, \eeq
where $|{\bf v}|$ is the speed of the star and $k$ describes the shape
of the velocity distribution near $\ve$.  Note that this approach is
only valid if the stellar velocities do indeed extend all the way to
$\ve$. If there is any truncation in the velocities then this approach
will underestimate the true $\ve$. 

Under the assumption that the Jeans theorem can be applied to the the
system, Equation (\ref{eq:ltdist}) can be understood by considering
the distribution function for the energies of the stars, $\epsilon$.
Assuming there is no anisotropy in the velocities, we can write the
asymptotic form of the distribution function as a power-law (K96),
\beq
\label{eq:df}
f(\epsilon)\propto\epsilon^k, ~~~{\rm where}~\epsilon=-(\Phi+|{\bf v}|^2/2),
\eeq
where $\Phi$ corresponds to the potential energy and $|{\bf v}|^2/2$ to the
kinetic energy. Again $k$ describes the shape of the velocity
distribution near $\ve$. Clearly $\Phi=-\ve^2/2$, which 
results in the following simple form for the asymptotic behaviour of
the velocity distribution,
\beq
\label{eq:vdist}
f(|{\bf v}|\,|\,\ve,k)\propto(\ve^2-|{\bf v}|^2)^k=[(\ve-|{\bf v}|)(\ve+|{\bf v}|)]^k.
\eeq

It could be argued that the velocity distribution near $\ve$ will not
follow the form given in equation (\ref{eq:vdist}) since the orbital
periods of such high velocity stars can be comparable to or larger
than the age of the system (hence invalidating Jeans Theorem). In this
case the velocity distribution near $\ve$ would be a superposition of
a small number of streams. This is likely to be an important issue
only in the tail of the velocity distribution (i.e. for stars with $v
\ga 3 \sigma$, where $\sigma$ is the velocity dispersion of the
system, see Helmi, White \& Springel [2002]). With this important caveat,
in this paper we shall follow the traditional assumptions reflected in
the above equations, and highlight possible limitations where
appropriate.

Intuitively we expect that $k>0$, in which case $f(|{\bf v}|) \rightarrow 0$
as $|{\bf v}| \rightarrow \ve$. However this is not a necessary condition
provided one accepts the presence of a discontinuity at
$f(|{\bf v}|=\ve)$. The likelihood of such distributions can be
observationally constrained (as will be shown in Section
\ref{sec:results}, we find that values of $k<0$ are strongly disfavoured).

Equation (\ref{eq:vdist}) can be further simplified in the limit of
($|{\bf v}| \rightarrow \ve$) to (\ref{eq:ltdist}) by neglecting the
$(\ve+|{\bf v}|)^k$ term, which has a strong systematic variation with
$k$.
However, throughout this paper, unless explicitly stated otherwise, we
adopt (\ref{eq:vdist}).

In LT90, it was argued that radial velocities alone provided the most
reliable constraints on $\ve$, since tangential velocities are much
more uncertain due to inaccuracies in measuring proper motions and
estimating distances. For example, a proper motion survey with typical
errors of $\sim10$ per cent in both proper motion and distance would
result in an error of $\sim60\kms$ for a star with velocity of $400\kms$,
whereas radial velocities can be measured with an accuracy of
typically less than a few $\kms$ (see Section \ref{sec:data}). 
In addition, since our work is motivated by the
current advances in radial velocity surveys, we shall not incorporate
the tangential velocities into our analysis.

To apply equation (\ref{eq:vdist}) to a sample of radial velocities, we
must average over the unknown tangential velocities,
\beq
\label{eq:tan_av}
f_{\rm r}(\vr|\ve,k)=\int f(|{\bf v}|\,|\,\ve,k)
\delta(\vr-\mathbf{v}.\mathbf{\hat{n}}) \mathrm{d}{\bf v},
\eeq
where $\mathbf{\hat{n}}$ is the unit vector along the line-of-sight.
Clearly, unless the lines-of-sight are isotropically distributed, this
equation is only valid for an isotropic distribution function. For our
dataset (the RAVE survey) we do not have all-sky coverage as RAVE only
monitors the Southern hemisphere. 
However, even if the distribution function is anisotropic, this
equation is still valid for a half-sky survey provided the mean motion
is small; as we shall see in Section \ref{sec:vmin}, we are dealing
almost exclusively with halo stars whose mean motion is indeed
small. Although the RAVE survey does not cover the entire Southern
sky, any corresponding bias should be negligible compared to the
relatively large statistical uncertainties in our final result. In any
case, if the distribution function is assumed to be isotropic then
the issue of sky coverage is immaterial since equation
(\ref{eq:tan_av}) is then valid for any sky coverage.

Evaluating equation (\ref{eq:tan_av}) for the LT90 formalism (equation
\ref{eq:ltdist}), we obtain,
\beq
\label{eq:vrdistlt}
f_{\rm r}(\vr|\ve,k) \propto (\ve-\vr)^{k+1}.
\eeq
For the formalism given in equation (\ref{eq:vdist}),
we integrate using cylindrical polar coordinates to obtain,
\beq
\label{eq:vrdist}
f_{\rm r}(\vr|\ve,k) \propto (\ve^2-\vr^2)^{k+1}.
\eeq

In order to constrain $\ve$ and $k$ for a sample of $n$ stars, we
employ the maximum likelihood method. The likelihood function
$l(\ve,k)$ for the unknown quantities to be estimated can be defined
as:
\begin{equation}
l \left (\ve,k \right) = \prod^n_{i=1} f_{\rm r} \left(v_{{\rm r},i}|\ve,k
 \right).
\label{eq:lik}
\end{equation}
where $f_{\rm r}(v_{{\rm r},i}|\ve,k)$ is given by either equation
(\ref{eq:vrdistlt}) or (\ref{eq:vrdist})
and $v_{{\rm r},i}$ are the radial velocities of the individual $n$ stars.

It can also be useful and sometimes more robust (especially for small
samples) to apply prior knowledge about $\ve$ and $k$ by way of
Bayes' Theorem.
Given $v_{{\rm r},i}$ radial velocity observations, the probability of
finding $\ve$ and $k$ in the ranges $\ve$ to $\ve + d\ve$ and $k$ to
$k + dk$, respectively, is given as:
\beq
\label{eq:lhood}
P(\ve,k|v_{{\rm r},i=1,...,n})=\frac{ P(\ve)P(k)\Pi^n_{i=1} P(v_{{\rm
r},i}|\ve,k)} {\int\int P(\ve')P(k')\Pi^n_{i=1} P(v_{{\rm
r},i}|\ve',k'){\rm d}\ve'{\rm d}k'}.   
\eeq
$P\left(v_{{\rm r},i}|\ve,k\right)$ is proportional to $f_{\rm
 r}(v_{{\rm r},i}|\ve,k)$.
$P(\ve)$ and $P(k)$ are the {\em a priori} probabilities (i.e., prior
knowledge) of $\ve$ and $k$, respectively. It is this equation, known
as the posterior distribution, with chosen reference priors $P(\ve)$
and $P(k)$, that will be maximized to determine $\ve$ and $k$.

In general the form of the distribution function given above (equation
\ref{eq:df}) is only true asymptotically  as $v \rightarrow \ve$, and
so to evaluate this probability we must impose a lower limit ($\vmin$)
on the magnitude of the radial velocities that we will consider in our
analysis; the choice of $\vmin$ is investigated further in Section
\ref{sec:vmin}.
As will be shown later in Section \ref{sec:data}, this form
provides a good fit to the data over our chosen range of velocities.

Once the distributions given in equations (\ref{eq:vrdistlt})
\& (\ref{eq:vrdist}) have been normalized such that
$\int_{\vmin}^{\ve} P(\vr|\ve,k) {\rm d}\vr = 1$
we can then evaluate equation (\ref{eq:lhood}).
One important factor in the evaluation of equation (\ref{eq:lhood}) is
the choice of $a priori$ probability; this is discussed in Section
\ref{sec:prior}.

\subsection{The bootstrap}
\label{sec:boot}

A short-coming of the method described above is that it assumes that
the distribution of stellar velocities is in equilibrium and
steady-state. However, there are many potential limitations to this
assumption, for example velocity substructure, binary systems for
which the centre-of-mass velocity has not been measured accurately or
non-equilibrium stars such as those ejected from binary systems
(including hyper-velocity stars, which are plausibly ejected from the
central regions of the Galaxy after interaction with the super-massive
black hole at the Galactic Centre; e.g. Brown et al. [2006]). All of
these mechanisms for potential contamination can statistically perturb
our underlying distribution function. They can change how the overall
observed distribution function is populated, (e.g. fluctuations due to
the addition of some orbital velocity with amplitude dependent on
phase and inclination of the binary orbit).  However, these mechanisms
are not well modelled, which makes it difficult to accommodate these
effects. It is crucial therefore to employ a bootstrap method, which
performs `resampling' on our original data-set to assess the
sensitivity to possible non-equilibrium stellar velocities. 

In brief, the bootstrap method involves randomly resampling the original
dataset (with replacement) to create artificial `bootstrap'
samples. Each bootstrap (re)sample provides one value each for $k$ and
$\ve$ that has maximum likelihood for that (re)sample of the data.
This is repeated a large number of times (typically 5000), and the
distribution of these maximum likelihood pairs is defined as the
`bootstrap distribution'.
The bootstrap distribution can be used to estimate the sampling
distribution of the maximum likelihood values for $k$ and $\ve$, e.g. as
obtained in equation (\ref{eq:lhood}). This is extremely useful,
because the fact we have a small sample means we cannot necessarily
rely on maximum likelihood estimators having converged to normality.
Therefore, we can use the bootstrap distribution as a means to model
the estimator sampling distributions.

Typically the bootstrap distribution is approximately normal, which
allows us to rely less on hope that the original sample size is large
enough for the central limit theorem to apply.   The bootstrap
distribution has bias if its mean values for $k$ and $\ve$ are not the
same as those found for the original sample.
Bias and skewness in the
bootstrap distribution are statistical signatures that can give us
a general understanding of possible fluctuations due to any
`contaminating' velocities, as explained above.  These signatures can
be studied using standard techniques (Davison \& Hinkley
1997). However, if the bootstrap distribution significantly deviates from
a normal distribution, steps must be taken to model the distribution.
The bootstrap standard error is the standard deviation of the
bootstrap distribution and the 90th-percentile confidence intervals
correspond to the regions containing 90 per cent of the maximum
likelihood pairs ($k$, $\ve$). This is useful, because now the bootstrap 
estimates of bias, standard error and confidence interval endpoints
are random variables. Their variances can be reduced by increasing the
number of bootstrap samples.  However, the quality of the bootstrap
approximation still depends on the original sample size.

\subsection{Considering the {\em a priori} probability for $k$ and $\ve$}
\label{sec:prior}

To evaluate equation (\ref{eq:lhood}) we must first choose the form of
the priors for $k$ and $\ve$. In previous work, the prior for $\ve$
was chosen to be $P(\ve)\propto 1/\ve$, since this is appropriate
for a variable that varies continuously from 0 to infinity  (Kendall
\& Stuart 1977). However, the choice of the prior for $k$ requires
more thought. In Section \ref{sec:formal} we expressed the velocity
distribution near the escape velocity assuming that the asymptotic
behaviour follows a power law (equations
\ref{eq:df}--\ref{eq:vdist}). Unfortunately, the range of feasible
values that one would expect for this exponent $k$ is
uncertain. Previous work has shown that to obtain meaningful limits on
$\ve$ from a single maximum likelihood analysis of a solar
neighbourhood sample  of stars one must assume some prior for $k$,
since only very large samples of stars would allow one to constrain
both $k$ and $\ve$ simultaneously. LT90 used Monte Carlo
simulations to estimate that samples of $\sim200$ stars with accurate
radial velocities will be required to constrain simultaneously $k$ and
$\ve$ for $v_{\rm min}=250\kms$ using the formalism given in equation
(\ref{eq:vrdistlt}).

For a given distribution function it is possible to predict the
behaviour of $k$. For example, a Plummer model (1911) has the
distribution function, $f(\epsilon)\propto\epsilon^{3.5}$,
i.e. $k=3.5$. Similarly, to first order approximation a Hernquist
(1990) model has a distribution function with $k=2.5$. However, LT90
argued that for a sample of stars free from selection effects, $k$
should lie between 1 and 2. They noted that an isolated system that
has undergone violent relaxation should have $k=1.5$ (Aguilar \& White
1986; Jaffe 1987; Tremaine 1987) and a collisionally relaxed system,
such as a globular cluster, should have $k=1$ (Spitzer \& Shapiro
1972). K96 chose to assume a uniform prior on $k\in[0.5,2.5]$, which
brackets this violent-relaxation value. Another approach one can
employ to understand the nature of $k$ is to use cosmological
simulations; we implement this approach in Section \ref{sec:abadi}.

Unlike previous work, we will also apply the bootstrap technique
(Section \ref{sec:boot}) to ascertain whether our results are
sensitive to possible problems with the data and allow confidence
estimation for the maximum likelihood values of $\ve$ and $k$. Since
we adopt bootstrap resampling  we cannot simply adopt the reference
priors chosen by LT90, without further investigation. In Appendix
\ref{appendix} we investigate the choice of prior by applying the
bootstrap analysis to simulated datasets. From this work we conclude
that a prior $P(k)\propto \sqrt{k}/(k+2),\, P(\ve)\propto
1/(\ve-\vmin)$ is the optimal choice, although for comparison we also
evaluate the bootstrap constraints using the classical LT90 choice,
i.e. $P(k)\propto1,\, P(\ve)\propto1/\ve$.

\section{Exploiting cosmological simulations to assess analysis techniques}
\label{sec:assess}

In this section we assess the reliability of the above techniques  in
recovering the escape velocity. We do this by analysing a suite of
four cosmological simulations. The first subsection deals with placing
{\em a priori} limits on the possible values of the exponent $k$,
which parametrizes the shape of the velocity distribution, while the
second subsection  deals with the effect of thin- and thick-disc
contamination on the recovery of $\ve$ and the choice of $\vmin$.

\subsection{The shape of the velocity distribution and its relation to
the parameter $k$}
\label{sec:abadi}

In Section \ref{sec:prior} we discussed the nature of the {\em a
priori} information that we can incorporate into the evaluation of the
likelihood function (equation \ref{eq:lhood}). We explained the
motivation behind the choice of prior in K96, i.e. a uniform prior on
$k\in[0.5,2.5]$. However, with cosmological simulations it is possible
to re-assess this choice. It was also noted in LT90 that if we have a
system where the stellar velocities do not extend up to $\ve$, the
statistical arguments of Section \ref{sec:formal} will always
underestimate the true $\ve$. So, in order to evaluate the ability of
the above statistical method to recover $\ve$, it is vital that we
understand the distribution of stellar velocities.

To test these issues we will analyse a set of four galaxies from
cosmological simulations, labelled KIA1--KIA4 (Abadi, Navarro \&
Steinmetz 2006). These galaxies have been introduced in earlier papers
and we advise interested readers to consult the following references
for details regarding the code and the numerical setup: Steinmetz \&
Navarro (2002), Abadi et al. (2003a,b), Meza et al. (2003, 2005).
These simulations, which are carried out  in a $\Lambda$CDM universe,
include the gravitational effects of dark matter, gas and stars, and
follow the hydrodynamical evolution of the gaseous component using the
Smooth Particle Hydrodynamics technique (Steinmetz 1996). The
following cosmological parameters were adopted for the $\Lambda$CDM
scenario: $H_0=65$ km s$^{-1}$ Mpc$^{-1}$, $\sigma_8=0.9$,
$\Omega_{\Lambda}=0.7$, $\Omega_{\rm CDM}=0.255$, $\Omega_{\rm
bar}=0.045$, with no tilt in the primordial power spectrum. All
simulations started at redshift $z_{\rm init}=50$, have force
resolution of order $1$ kpc, and mass resolution so that each galaxy
is represented, at $z=0$, with $\sim 125,000$ star particles. The
range of masses of the stellar particles in these four simulations at
$z=0$ is $10^5 - 10^7 {\rm M}_\odot$.  Readers who wish to find a
general overview of these simulations are recommended to consult
section 2 of Abadi et al. (2006), while a more detailed description
can be found in the references given above.

There is various evidence that demonstrates the reliability of these
simulations. Section 3.2 of Abadi et al. (2003a) discusses the
photometric properties. Here they note the good agreement between
their surface brightness profile and a superposition of an $R^{1/4}$
spheroid plus exponential disc. The issue of the outer parts of the
halos of these simulated galaxies are discussed in Abadi et
al. (2006). The mass distribution in these outer parts is seen to be
consistent with the distributions of Galactic and M31 globular
clusters. In addition, the surface brightness in these outer regions
is well fit by a Sersic law and is consistent with observations of
both the halos of M31 and the Milky Way. Support that the kinematics
of these halos are reliable can be drawn from the fact that they are
consistent with the most recent observations of halo tracers
(e.g. Battaglia et al. 2005, 2006).

In order to allow for a fair comparison between these four galaxies
and the Milky Way, we rescaled each of them so that their circular
velocity at their virial radius\footnote{We define the virial radius
as the radius at which the mean density is 200 times the critical
density for closure ($\rho_{\rm cr}=3H^2/8\pi
G\approx7.9\times10^{-27}\,{\rm kg\,m^{-3}}$ for $H=65 \kms\,{\rm
Mpc}^{-1}$).} is equal to $140 \kms$. This value of $140\kms$ is a
typical value one could expect from using an NFW profile to
extrapolate the circular velocity at the solar neighbourhood, $\sim
220$~km/s, to the virial radius. This results in a small rescaling
factor ($\gamma$) of between 0.9--1.3, where distances, velocities and
masses are rescaled by $1/\gamma$, $1/\gamma$ and $1/\gamma^3$,
respectively. The properties of our rescaled galaxies are given in
Table \ref{tab:KIAprop}.

For these simulated galaxies we need to estimate the escape velocity
at the solar radius (which, for these simulations, we take to be
8.5~kpc from the centre of the model galaxy ). To do this we define
the escape velocity as the velocity required to get to $3{\rm
r_{vir}}$, i.e. we define the zero of the gravitational potential to
be at this radius. Although this choice of $3{\rm r_{vir}}$ is somewhat
arbitrary, it is  comparable to the separation of the Milky Way and
its nearest massive neighbour M31. We then determine the potential at
the solar radius to deduce the escape velocity.
The resulting values for $\ve$ are
between 581 and 653 $\kms$ (see Table \ref{tab:KIAprop}). The stellar
particles contribute a significant amount of mass, namely between 60
and 70 per cent of the total mass interior to the solar circle.

\begin{table}
\begin{tabular}{lcccc}
\hline Name & Rescaling & $M_{\rm vir}$ & $r_{\rm vir}$ & $\ve$\\ &
Factor & ($10^{10} {\rm M}_\odot$) & (kpc) & ($\kms$)\\ \hline KIA1 &
1.30 & 120.59 & 230.73 & 581.06\\ KIA2 & 0.88 & 127.64 & 235.15 &
640.17\\ KIA3 & 0.87 & 129.87 & 236.50 & 652.62\\ KIA4 & 1.15 & 126.87
& 234.67 & 582.92\\ \hline
\end{tabular}
\caption{The properties of our simulated galaxies used in Section
\ref{sec:abadi}. The second column shows the rescaling factor that has
been applied, which is chosen so as to enforce the circular velocity
at the virial radius to be $140 \kms$. The second and third columns
refer to the virial mass and radius of the rescaled galaxies,
respectively. The final column gives the escape velocity at the solar
radius (8.5 kpc), as calculated according to the prescription
described in Section \ref{sec:abadi}.}
\label{tab:KIAprop}
\end{table}

Given these four simulated galaxies, the first question which needs to
be addressed is whether the velocity distribution of the stars differs
from that of the dark matter particles, and in particular whether the
stellar velocity distribution extends all the way to $\ve$.
The stellar distribution must differ from that of the dark-matter
particles since the two populations have different density
distributions.

We investigate this issue by plotting the velocity distributions for the
stellar and dark matter components in our four galaxies. Since the
disc component of these simulations do not always provide a wholly
accurate representation of the Milky Way thin and thick discs, we only
plot the distribution of stellar particles defined as belonging to the
`spheroid' population, according to the decomposition of Abadi et
al. (2003b).  This decomposition is based on the particles' orbital
parameters and results in three populations: a halo with little net
rotation, a thin disc with stars on nearly circular orbits and a thick
disc with properties intermediate between the other two
components. The halo component comprises between 10 and 40 per cent
of the total number of stellar particles inside the virial radius.

\begin{figure}
\plotone{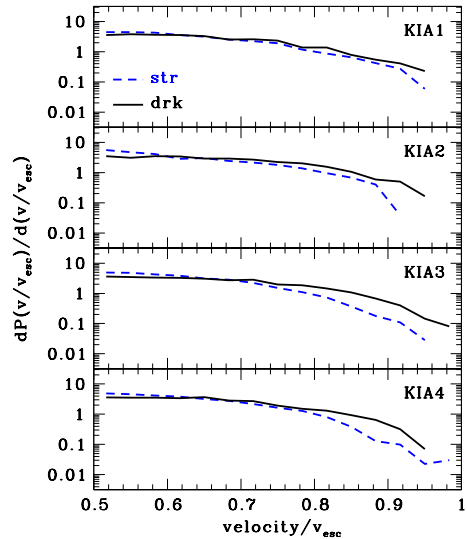}
\caption{This figure shows the velocity distribution of the stars
(dashed) and dark matter particles (solid) from our four simulated
galaxies. These distributions are for particles whose distance from
the centre of the galaxy is between 3 and 14 kpc, and the stellar
particles are those belonging to the `spheroid' population.}
\label{fig:vel_dist}
\end{figure}

Our exclusion of the disc populations from these simulations can be
further justified when one considers that we expect our high velocity
sample of stars to contain only limited contamination from the Milky
Way thick disc and negligible contamination from the thin disc (see
Section \ref{sec:vmin}). These velocity distributions are calculated
for particles whose distance from the centre of the galaxy is between
3 and 14 kpc. Since the escape velocity varies over this large range
of radii, we rescale each velocity by the mean escape velocity at that
radius. The resulting distributions are shown in
Fig. \ref{fig:vel_dist}. It can easily be seen from this figure that
the velocity distributions differ between the stellar and dark matter
components, with the stellar particles' velocities concentrated
towards smaller values.  This is due to the difference between the
density profiles of the stellar halo and dark components, since the
stellar particles are more concentrated and hence have smaller
velocities.  At the solar radius, the respective density distributions
have power-law slopes of $\sim-2$ for the dark matter and $\sim-3$ for
the stellar halo, which is consistent with what is found for our
Galaxy (e.g. Morrison et al. 2000).  Despite the fact that the stellar
velocities are smaller, it is encouraging to see that these velocities
do indeed extend all almost the way to $\ve$.
This is vital if we are
to use this statistical analysis described in Section
\ref{sec:analysis} to determine $\ve$ for the Milky Way. Although the
decline in the velocity distribution for values close to $\ve$ appears
to be sharper for the stellar particles, 
all of our galaxies have
stellar particles with velocities greater than $0.9\ve$. As a
consequence, even if there is a truncation beyond $0.9\ve$, the effect
on $\ve$ cannot be more than 10 per cent.

Since these simulations indicate that the velocity distribution of
stars clearly differs from that of the dark matter particles, one
needs to ask how this affects the exponent $k$. To investigate this we
evaluate the likelihood function (equation \ref{eq:lhood}) for samples
of stellar and dark particles, but keeping $\ve$ fixed at the true
value calculated above, i.e., \beq
\label{eq:fixedve}
P(k|v_{{\rm r},i=1,...,n})=\frac{ P(k)\Pi^n_{i=1} P(v_{{\rm r},i}|k)}
{\int P(k')\Pi^n_{i=1} P(v_{{\rm r},i}| k'){\rm d}k'}, \eeq where we
have assumed a uniform prior on $k$.

To evaluate equation (\ref{eq:fixedve}) we need to construct a sample
of particles equivalent to our observed solar neighbourhood
stars. This is done by calculating the radial velocities of particles
contained within a series (in azimuth around the galaxy) of
non-overlapping spheres of radius 2 kpc located at a distance of 8.5
kpc from the Galactic centre. These spheres are all chosen to lie in
the plane of the Galaxy. Since the presence of significant mean
orbital rotation in our sample of stars will invalidate the
assumptions made in Section \ref{sec:analysis}, we only use the
stellar particles defined as belonging to the non-rotating `spheroid'
population, according to the decomposition of Abadi et al. (2003b).
In order to increase our statistics we also include spheres of radius
1 and 3 kpc located at radii 4.25 and 12.75 kpc; we ensured that there
is no trend between the exponent $k$ and the radial location of the
spheres within our range of values. To combine spheres located at
different radii, the velocities of particles within each sphere are
rescaled by the escape velocity at the centre of the sphere.

\begin{figure}
\plotone{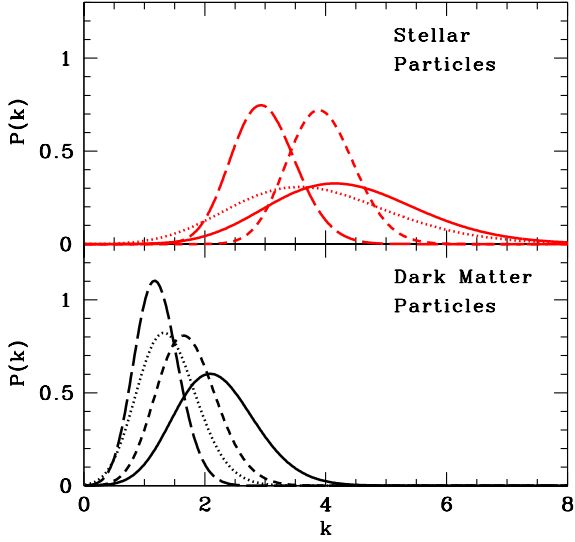}
\caption{This figure shows the likelihood estimates for the exponent
$k$, which denotes the shape of the velocity distribution. This is
shown for both the stellar (thick) and dark matter (thin) high
velocity `solar neighbourhood' samples. The solid, dotted, short
dashed and long dashed lines represent simulated galaxies KIA1, KIA2,
KIA3 and KIA4, respectively.}
\label{fig:kprior}
\end{figure}

By evaluating equation (\ref{eq:fixedve}) we are able to deduce the
likelihood estimates for the value of $k$ for the stellar and dark
matter populations for each of our four simulated galaxies. The
resulting likelihood distributions of $k$  are shown in
Fig. \ref{fig:kprior}.

As can be seen from this figure, the value of $k$ significantly
differs for the two populations. For these simulated galaxies we can
see that the dark matter particles appear to match the predictions
from the literature, i.e. $k \in [0.5,2.5]$ (see above). However, the
stellar sample does not appear to match this expectation; for these
$k$ is shifted towards significantly larger values. Since all attempts
to constrain $\ve$ must be based on stellar samples, it means that a
uniform prior on $k \in [0.5,2.5]$ will not provide an accurate
measure of the escape velocity. To estimate a better choice of prior
from these stellar likelihood intervals, we take the same range of $k$
but centre it on the mean likelihood of 3.7, i.e. $k \in [2.7,4.7]$.

At first sight it may appear from Fig. \ref{fig:kprior} that the
constraints on $k$ for galaxies KIA1 and KIA2 differ slightly from the
other two galaxies. However, this can be understood when one considers
the fact that KIA1 and KIA2 have fewer total numbers of stellar
particles than KIA3 or KIA4. As a consequence the constraints on $k$
appear to be shifted towards higher values for these galaxies. This is
simply due to the fact that when one has fewer particles, although the
regime $k\rightarrow0$ is still well constrained (since negative $k$
is very strongly disfavoured), it is harder to 
constrain large values of $k$. Given this fact, an alternative
approach to determine a prior could be to take the range covered by
the joint 90 per cent confidence interval from KIA3 and KIA4 alone. If
we do this we obtain a range $k \in [2.3,4.7]$, which is very similar
to that found above. Given the fact that there is very little
difference, we choose to adopt the former range (i.e. $k \in
[2.7,4.7]$) for the subsequent analysis.\footnote{If we repeat the
subsequent analysis using this prior, we obtain the following
constraints on $\ve$ $494\kms < \ve < 605 \kms$ (90 per cent
confidence), with a median likelihood of $540\kms$. These constraints
are practically indistinguishable from those found by adopting the
prior using all four galaxies (see Section \ref{sec:results}).}

It is important to check whether the results presented in
Fig. \ref{fig:kprior} are sensitive to our choice of rescaling, since
there are a variety of different rescalings that one can adopt. To
test this we adopt an alternative rescaling that enforces a circular
velocity of $220\kms$ at 8.5 kpc. This results in a rescaling factor
of 1.11, 0.78, 0.81, 0.83 for galaxies KIA1--KIA4, respectively, which
is not far removed from our original values (see Table
\ref{tab:KIAprop}). Reassuringly, when we repeat the above analysis
the resulting likelihood estimates for $k$ are wholly consistent for
each of our simulated galaxies.

\subsection{The choice of $\vmin$}
\label{sec:vmin}

\subsubsection{Estimating the level of contamination from the disc}

We wish to address the choice of $\vmin$, i.e. the minimum velocity
that is included in a sample of `high velocity' stars. For their
analysis of radial velocities, LT90 chose a value of $250\kms$, while
K96 chose $300\kms$, preferring to reduce any systematic errors at the
expense of increased statistical errors. When considering the choice
of $\vmin$ it is crucial to note that the analysis discussed in
Section \ref{sec:analysis} relies on the assumption that the
velocities are isotropic and there are no mean motions. This is
especially important because RAVE fields are not isotropically
distributed on the sky; if there is any net rotation then the
averaging over the unknown tangential velocities performed in equation
(\ref{eq:tan_av}) is in error.  While we might expect the
mean motion of the halo stars to be negligible compared to the
relatively large statistical errors that will be present in any
current analysis, the possible presence of the rotating thin- and
thick-disc population in any sample cannot be ignored.

In the absence of any additional information (such as metallicities,
etc), the best discriminator which can be applied to radial velocity
samples is one based on the radial velocity itself. Given the expected
small  dispersion in velocities of disc stars, one would hope that a
sufficiently large value of $\vmin$ will be able to filter out the
disc population. We will investigate this issue by first estimating
the fraction of disc stars that we may expect in our high velocity
sample. To do this we construct a simple toy model containing each of
the three components, i.e. the thin and thick discs and halo.

Before one can estimate the fraction of disc stars that may be present
in our high velocity sample, one first must ask what the total
fraction of disc stars will be over all velocities. From the
literature we find that in the solar neighbourhood the mass density of
the three components can be estimated as $3.8\times10^{-2}, \,
1.3\times10^{-3}$ and $1.5\times10^{-4}\,\,\sm\,{\rm pc}^{-3}$ for the
thin disc, thick disc and halo, respectively (Jahrei{\ss} \& Wielen
1997; Fuchs \& Jahrei{\ss} 1998; Ojha 2001). However, our sample of
stars will be drawn from a finite volume, and so we would expect to
find significantly more halo stars than this value of 0.4 per cent. To
obtain the final fraction we need to know how the mass density of the
components varies spatially and also the approximate boundary of our
sample volume. We take a value of 2 kpc for this latter quantity and
estimate the former by assuming that the mass density has negligible
radial dependence in the plane and is solely determined by the
vertical scale-height ( for which we adopt the values of  260 and 860
pc for the thin and thick discs, respectively, consistent with the
local normalisations; Ojha 2001). We assume that the halo density is
constant. These assumption should be valid provided our RAVE sample
volume is not too large. The final property that we include is that
our RAVE sample excludes stars within 25 degrees of the plane. If we
fold all this information into a toy model, then the resulting value
for the fractional contribution of the three components  before
considerations of kinematics  becomes 73.5, 19.5, and 7.0 per cent for
the thin disc, thick disc and halo, respectively. Clearly we cannot
calculate these fractions with any high degree of certainty since we
have not accounted for the luminosity function of the three components
and also the local mass densities are subject to much
discussion. However, we believe that these values provide a
reasonable estimate and should be suitable for our purposes.

The final ingredient necessary for calculating the expected level of
contamination from disc stars is to include the kinematics of the
three components. This is done by simply assuming Gaussian velocity
distributions with mean orbital rotation and dispersions as given in
Table \ref{tab:vel_disp}. The halo parameters are taken from Chiba \&
Beers (2000). Although their value of $\left<v_\phi\right>=40\kms$ is
in conflict with our assertion that the halo must have no net
rotation, this value is clearly small compared to the dispersions.
The thin- and thick-disc dispersions are deduced from the data of
Nordstr\"om et al. (2004) following the prescription of Binney \&
Merrifield (1998, section 10.4; see also Reddy, Lambert \& Allende
Prieto 2006); namely that thin disc stars satisfy age$<$10 Gyr, ${\rm
\left[Fe/H\right]>-0.4}$, while thick-disc stars satisfy age$>$10
Gyr, ${\rm \left[Fe/H\right]<-0.4}$. 

\begin{table}
\begin{tabular}{lcccc}
\hline Component & $\left<v_\phi\right>$ & $\sigma_R$ & $\sigma_\phi$
& $\sigma_z$\\ & ($\kms$) & ($\kms$) & ($\kms$) & ($\kms$) \\ \hline
Thin Disc & 209 & 29 & 18 & 14\\ Thick Disc & 174 & 68 & 55 & 38\\
Halo & 40 & 141 & 106 & 94\\ \hline
\end{tabular}
\caption{ Velocity dispersions of the three components in the toy
model of Section \ref{sec:vmin}. The halo dispersions are taken from
Chiba \& Beers (2000). The thin- and thick-disc dispersions are deduced
from the data of Nordstr\"om et al. (2004) following the prescription
of Binney \& Merrifield (1998; section 10.4); namely that thin disc
stars satisfy age$<$10 Gyr, ${\rm \left[Fe/H\right]>-0.4}$, while
thick-disc stars satisfy age$>$10 Gyr, ${\rm \left[Fe/H\right]<-0.4}$. 
We adopt a value of $220\kms$ for the local standard of rest and take
the solar motion from Dehnen \& Binney (1998).}
\label{tab:vel_disp}
\end{table}

\begin{figure}
\plotone{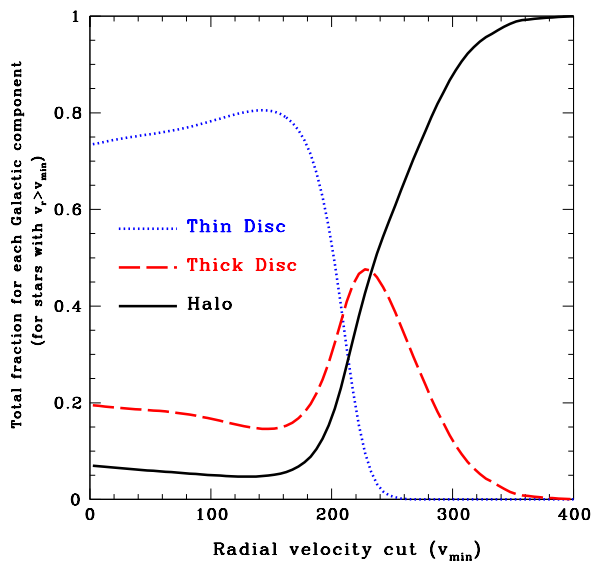}
\caption{This figure shows the fractional contribution of each
Galactic component as a function of $\vmin$, i.e. the minimum radial
velocity of the sample. These results come from the toy model
described in Section \ref{sec:vmin}.}
\label{fig:contam}
\end{figure}

Given this toy model we can now estimate the contribution of each
component in a sample of stars with radial velocity greater than some
cut-off $\vmin$. To do this we take each observed star in the RAVE
catalogue and generate 100 mock stars along this line-of-sight
according to our toy model. We can then calculate the fractional
contribution of each component as a function of $\vmin$, which is
shown in Fig. \ref{fig:contam}. From this figure it can be seen that
for $\vmin>250\kms$ the contamination from the thin disc is
negligible. However, thick-disc stars will be present for velocities
as high as $400\kms$. As stated above, K96 chose a value of
$\vmin=300\kms$, for which our toy model predicts a contamination of
approximately 10 per cent. The contamination rises to $\sim20$ per
cent if $\vmin=280\kms$ and $\sim40$ per cent if $\vmin=250\kms$. Although the
optimal value of $\vmin$ is dependent on the line-of-sight (for
example as a function of longitude), we follow previous studies and do
not adopt a spatially varying $\vmin$.

\subsubsection{Quantifying the effect of disc contamination}
\label{sec:quantify_contam}

We now need to test whether this level of contamination will affect
our results. To do this we return to the simulations of Abadi et
al. (2006) that were introduced in Section \ref{sec:abadi}. Since the
disc components in these galaxies do not always provide a wholly
accurate representation of the Milky Way disc, we instead include the
contamination from the thick disc according to the toy model described
above. We investigate two fiducial cases, namely thick-disc
contamination of 10 per cent and 20 per cent (note that for velocity
cuts of $\vmin\ge250\kms$ we predict that the contamination from the
thin disc will be negligible). For the halo component we take the
`spheroid' star particles as defined in Section \ref{sec:abadi}. In
addition, we restrict ourselves to the analysis of only two of our
simulated galaxies (KIA3 and KIA4) since they have significantly more
particles than the other two. In order to make a fair comparison
between the two galaxies we rescale the velocities of the halo
component so that $\ve=600\kms$ and ensure that each sample contains
100 stars. For the minimum velocity cut we take a value of
$\vmin=300\kms$.

Given these two samples of simulated stars with 10 and 20 per cent
thick-disc contamination, we evaluate equation (\ref{eq:lhood}). The
results from 
this analysis indicate that the presence of a thick-disc population
has a noticeable, but not hugely significant, effect. The peak of the
likelihood distribution is shifted towards larger $k$ and $\ve$,
although for a given value of $k$ the predicted value of $\ve$ is
lower. Therefore, when one applies the prior calculated in Section
\ref{sec:abadi} (i.e. $k\in[2.7, 4.7]$) the predicted escape velocity
is reduced. For a thick-disc contamination of 20 per cent, this shift
is approximately $-30\kms$ for both KIA3 and KIA4; for a contamination
of 10 per cent this shift is reduced to $-15\kms$ for KIA3 and
$-10\kms$ for KIA4.

In conclusion, we concur with K96 and propose that a value of
$\vmin=300\kms$ is a safe choice which should allow us to obtain a
reliable and robust determination of the escape velocity.

\section{Data}
\label{sec:data}

The following sections describe the data that we will use to constrain
the escape velocity.

\subsection{The RAVE project}
\label{sec:rave}

The Radial Velocity Experiment (RAVE) is an ambitious survey to
measure radial velocities and stellar atmosphere parameters 
(temperature, metallicity, surface gravity) of up to one million stars
using the 6dF multi-object spectrograph on the 1.2-m UK Schmidt
Telescope of the Anglo-Australian Observatory (AAO).  The RAVE survey
is ideal for constraining $\ve$ because it is a magnitude limited
survey ($9<I<12$), which means that it avoids any kinematical biases,
unlike catalogues constructed using high proper-motion stars. Given
the RAVE magnitude range, the catalogue will consist of giants up to a
distance of a few kpc and local dwarfs.  The project is described in
detail in the paper accompanying the first data release (Steinmetz et
al. 2006). It is foreseen that the RAVE project will run until
2010. This survey will represent a giant leap forward in our
understanding of our own Milky Way galaxy, providing a stellar
kinematic database larger than any other survey proposed for this
coming decade. For our analysis we use an internal data release
(Summer 2006) containing radial velocities for over 50,000 stars
covering an area of nearly 8000 square degrees.

\subsection{Our high radial velocity RAVE sample}
\label{sec:ravesample}

To construct a catalogue of high radial velocity objects we must first
convert our heliocentric velocities into Galactic standard of rest
frame velocities (also known as Galactocentric radial velocities; see
for example equation (10-8) of Binney \& Tremaine 1987). This is done
by taking the Local Standard of Rest to be $220\kms$ and the solar
motion to be $(10.00,5.25,7.17)\kms$ (Dehnen \& Binney 1998). Note
that unless stated otherwise, all velocities will be quoted in the
Galactic standard of rest frame. We then take all stars greater than
$\vmin = 300 \kms$.

It is vital to ensure that our sample is free from contamination from spurious
measurements. To this end we enforce a high threshold for the value of
the Tonry-Davis (Tonry \& Davis 1976) correlation coefficient between
the observed spectrum and the template spectrum ($R>10$). We perform
additional checks to verify the reliability of the data in the
following section. It should also be noted that currently the
metallicities and gravities have not been accurately determined for
the RAVE data, so we will only use the radial velocities in our
analysis of the escape velocity.

A total of 16 stars pass these criteria. Two of these have $10<R<15$,
but the velocities for both stars have been confirmed by follow-up
observations (see the following Section).

\subsection{Data validation}
\label{sec:validation}

This subsection discusses the methods that have been adopted to verify
the reliability of the RAVE radial velocity measurements of our sample
of 16 high velocity stars.  For a comprehensive discussion of the
quality of the RAVE data in general, we refer readers to the
data-release paper (Steinmetz et al. 2006).

\begin{figure}
\plotone{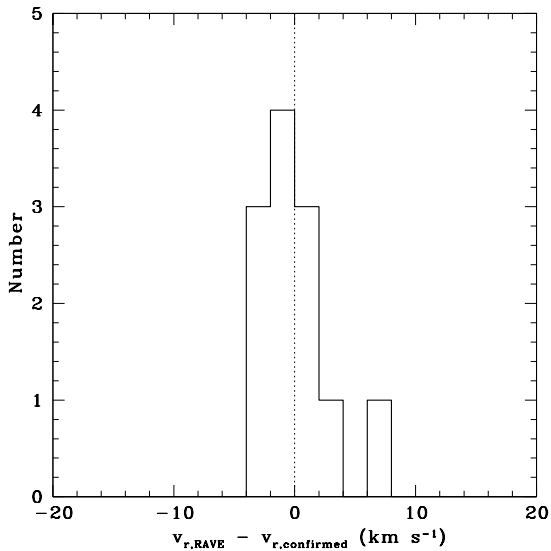}
\caption{Results from the follow-up work described in Section
\ref{sec:validation} for 12 of our high radial velocity RAVE
stars. The horizontal axis shows the difference between the velocity
as reported in the RAVE catalogue compared to the weighted mean of all
velocities taken during the follow-up campaign. Note the good
agreement between the two measurements. Typical errors are $\sim
2\kms$ for the RAVE catalogue and $<1 \kms$ for the follow-up
velocities.}
\label{fig:mary}
\end{figure}

We have obtained follow-up observations of a selection of RAVE stars
using the 2.3m Advanced Technology Telescope at Siding Springs
Observatory (Steinmetz et al. 2006). The data were taken using the
Double Beam Spectrograph, giving slit spectra between 8000 -- 8900
$\AA$ at a resolution similar to the RAVE observations (0.55 $\AA/{\rm
pix}$). In general, the radial velocity measurements show reasonable
agreement with the RAVE values to within $\sim3 \kms$. Given the fact
that we wish to use only the highest radial velocity stars for our
$\ve$ analysis, this level of discrepancy is negligible.  Among this
selection of re-observed stars there were 12 of our high radial
velocity RAVE stars, all of which show good agreement.  A smaller
subsample of RAVE stars were also observed with the echelle
spectrograph on the 3.5m Telescope at Apache Point Observatory (APO),
New Mexico. The single-slit echelle has a spectral resolution of
37,000 and covers the entire optical wavelength range (3500$\AA$-1$\mu
m$). Nine of the observed stars were high-velocity stars.  Similarly
to the 2.3m data, the uncertainties are dominated by RAVE errors, and
the radial velocity measurements agreed with the RAVE velocities
within 3$\sigma$.  In Fig. \ref{fig:mary} we show the results of the
follow-up observations for these 11 stars. All show good agreement,
with none having a discrepancy of more than $7 \kms$.

Although we found good agreement for our follow-up velocities, there
were two exceptions not mentioned above. Two of our stars showed
behaviour consistent with being binary stars. One of these stars
(C2129509-080418) has 4 observations taken over a period of 18 months,
while another (C1437570-280154) has 2 observations taken over a period
of 26 months. From our measured velocities we can place a lower limit
on the semi-amplitude of the velocity variations and we find values of
$10.6~\kms$ and $15.4~\kms$ for C1437570-280154 and C2129509-080418,
respectively. These values are typical of single-lined spectroscopic
binaries in Latham et al.~(2002), for which 68 per cent have
semi-amplitudes within the range $3~\kms$ to $15~\kms$. Therefore if
we assume that the actual semi-amplitude for our two RAVE stars is
close to our observed lower limit, we can estimate the centre of mass
velocity by averaging the highest and lowest velocities for each
star. Given the statistics of the Latham et al. (2002) sample, in
particular that less than 10 per cent of their binaries have
semi-amplitude greater than $20\kms$, we can predict that the probable
error for the estimated velocity of our RAVE stars is less than
$10~\kms$. Both of these centre of mass velocities are above $\vmin$
and hence we retain these for our analysis. Our fraction of binary
stars (2 out of 16) is comparable to the fraction of single-lined
spectroscopic binaries found in the Latham et al. (2002) sample (171
out of 1359).

In summary, out of our final sample of 16 RAVE stars with
Galactocentric velocity greater than $300\kms$, 14 have repeat
observations (including the two binary stars). The total number of
observations for each star varies from between one and four (see Table
\ref{tab:rave}).

\subsection{The final high-velocity RAVE sample}
\label{sec:ravefin}

\begin{figure}
\plotone{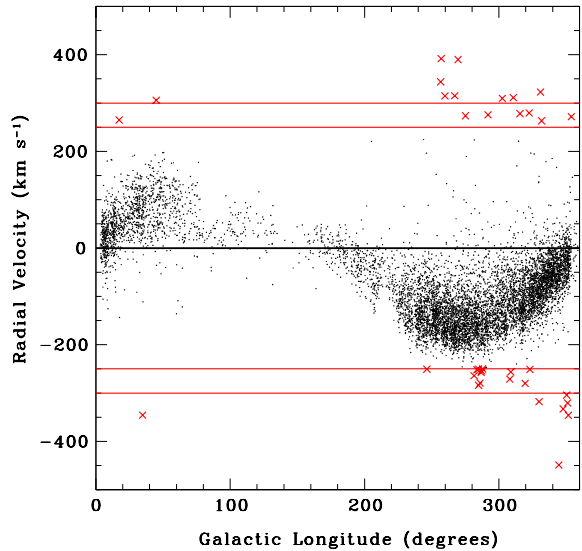}
\caption{The relation between radial velocity (corrected for Solar
motion) and longitude for stars in the RAVE catalogue. Note that the
signature of the disc is clearly visible. The horizontal lines
correspond to $\vr=-300,-250,0,+250,+300\kms$. The crosses simply
denote stars with $\left|\vr\right|>250\kms$.}
\label{fig:checkrot}
\end{figure}

Given this high quality RAVE data, we are now able to construct a
final catalogue of high radial velocity stars. Since many of our stars
now have repeat observations, we choose to adopt the weighted mean of
all measurements as our definitive velocity, with the exception of the
two binary stars for which we give our estimate of the center of
binary mass motion.  These are tabulated in Table \ref{tab:rave} and
the velocity distribution is shown in the inset of
Fig. \ref{fig:data}. In Fig. \ref{fig:checkrot} we show how the radial
velocities vary as a function of Galactic longitude. This plot clearly
shows the signature of the Galactic disc and from this one can obtain
an understanding of why a value of $\vmin\approx250\kms$ results in
significant contamination from the disc; if the mean rotational
velocity of our sample is close to zero (as we would like for a halo
population), then there should be an equal number of stars with
positive and negative radial velocity for a given longitude. However,
for $l\approx270$ there is clearly a greater number of stars with
radial velocities in the range $\vr\in(-300,-250)$ compared to
$\vr\in(250,300)$, indicating contamination by a rotating
component. Note that this asymmetry is not evident for stars with
$\left|\vr\right|>300\kms$, which supports our choice of
$\vmin=300\kms$.

\begin{table*}
\begin{tabular}{lcccccccc}
\hline Name & $\vr$ & $\delta\vr$ & RA (J2000) & Dec (J2000) & $\ell$
& b & $I$ & Total No. of\\ & ($\kms$) & ($\kms$) & ($^\circ$) &
($^\circ$) & ($^\circ$) & ($^\circ$) & (mag) & Observations \\ \hline
C1012254-203007 & 314.7 & 0.5 & 153.10625 & -20.50194 & 259.83 & 28.76
& 12.2& 3 \\ C1022369-140345 & 391.8 & 0.5 & 155.65408 & -14.06267 &
257.04 & 35.23 & 12.0& 2 \\ C1032508-244851 & 315.3 & 1.4 & 158.21199
& -24.81439 & 267.08 & 28.24 & 11.9& 1 \\
C1100242-024226 & 344.0 & 0.7 & 165.10114 & -2.70722 & 256.67 & 49.92
& 11.7& 4 \\
C1250398-030748 & 309.8 & 1.1 & 192.66612 & -3.13019 & 302.55 & 59.74
& 12.1& 2 \\
C1437570-280154 & -317.6 & --- & 219.48779 & -28.03169 & 329.89 &
29.21 & 11.3& 2 \\ C1508217-085010 & -304.3 & 1.3 & 227.09082 &
-8.83622 & 350.44 & 41.07 & 11.3& 2\\ C1519196-191359 & -448.8 & 0.7 &
229.83192 & -19.23309 & 344.64 & 31.41 & 12.1& 4 \\ C1536201-144228 &
-345.9 & 1.0 & 234.08397 & -14.70792 & 351.75 & 32.12 & 12.4& 3 \\
C2041305-113156 & -345.8 & 0.4 & 310.37735 & -11.53244 & 34.81 &
-29.61 & 11.8 & 3 \\ C2129509-080418 & 305.91 & --- & 322.4620  &
-8.0717 & 44.9807 & -38.7363 & 12.5 & 4 \\
C2214430-480306 & -332.9 & 1.8 & 333.6800 & -48.0519 & 347.71 & -53.16
& 11.7 & 1 \\
T4931\_00266\_1 & 390.0 & 0.4 & 175.27388 & -1.54536 & 269.56 & 56.70
& 10.6& 4 \\
T7524\_00065\_1 & 322.7 & 0.8 & 4.61800 & -39.00944 & 330.93 & -76.27
& 9.8& 3 \\ T7535\_00160\_1 & 311.3 & 0.8 & 10.37617 & -40.98189 &
310.70 & -76.00 & 10.2& 3 \\ T8395\_01513\_1 & -320.5 & 0.9 &
300.80170 & -48.10783 & 351.10 & -31.50 & 10.0& 3 \\
\hline
\end{tabular}
\caption{The high velocity sample of RAVE stars. See Section
\ref{sec:data} for details about these stars. The identifiers are
taken from the RAVE input catalogue, while the $I$-band magnitudes are
from the DENIS catalogue (Epchtein et al. 1997). Note that the
magnitudes of three stars, C1437570-280154, C2129509-080418 and
T7524\_00065\_1, are not available in the DENIS catalogue and so the
values for these stars are taken from USNO-B (Monet et
al. 2003). Column one shows the Galactocentric radial velocity and
column two shows the error in the velocity; where we have more than
one observation for a star we adopt the weighted mean. Two stars in
our sample are believed to be binaries (C1437570-280154 and
C2129509-080418) and for these we give an estimate of the
center of binary mass motion (see Section \ref{sec:validation}).}
\label{tab:rave}
\end{table*}

\subsection{Augmenting our high velocity sample with stars from
archival surveys}
\label{sec:include}

Since we would like our sample of stars to be as large as possible, we
incorporate additional stars from the Beers et al. (2000) catalogue of
metal poor stars. It is important to note that this sample is
kinematically unbiased, which is important if we are to combine
datasets in this way. This sample is ideal since it contains metal
poor stars, which are preferentially halo stars.
The Beers et al. (2000) sample provides a total of 17 stars faster
than $300\kms$, once we have removed three stars for which the
distance estimate indicates that they are further than 5 kpc away (all
of the retained stars have distances of less than 2.5 kpc). These
archival stars are given in Table \ref{tab:beers}.

This brings the total number of stars in our full augmented sample to
33, which is a significant improvement on the number of stars used in
LT90 (15 stars with $\vr>250\kms$) and K96 (10 stars with
$\vr>300\kms$).

\begin{figure}
\plotone{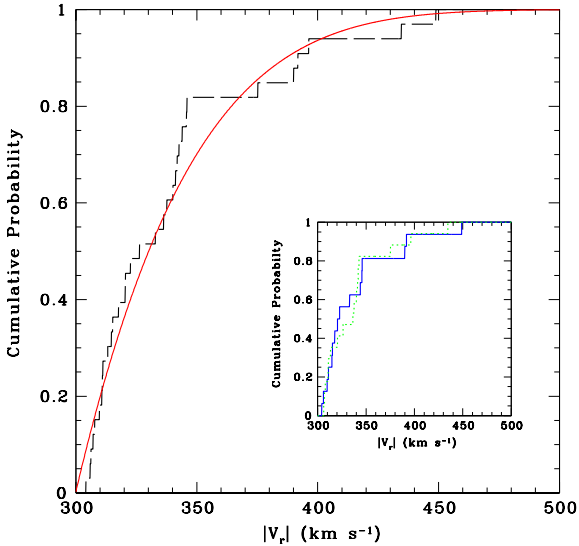}
\caption{The cumulative distribution for the magnitude of the radial
velocities. The dashed 
line shows the velocities for our sample of high radial velocity
stars, which has been constructed by augmenting a sample of RAVE stars
with a supplementary archival survey (Beers et al. (2000); see Section
\ref{sec:data}). The solid line corresponds to the maximum likelihood
evaluated in Section \ref{sec:results} ($\ve=616\kms$, $k=7.1$). The
inset shows the comparison between the 16 RAVE stars (solid) and the
17 archival stars (dotted), showing no clear discrepancy between these
two samples.}
\label{fig:data}
\end{figure}

The velocity distribution for this larger augmented sample is shown in
Fig. \ref{fig:data}. Note that the inset of this figure compares the
distribution of RAVE stars with the distribution of our archival
stars from Beers et al. (2000).
The Kolmogorov-Smirnov test indicates no significant discrepancy
between these two distributions (14 per cent probability that they
come from different distributions), so there is no inconsistency in
combining the two data sets.
In addition, similar to the RAVE sample (as was shown in
Fig. \ref{fig:checkrot}), we reassuringly find no correlation between
radial velocity and Galactic longitude.
In Section \ref{sec:results} we check that the process of combining
datasets does not introduce any obvious bias by carrying out the
likelihood analysis on both the combined sample as well as a sample
consisting solely of our 16 RAVE stars.

\section{Results}
\label{sec:results}

We now wish to use the sample of high velocity stars described above
to constrain the local escape velocity. To do this we apply the
techniques outlined in Section \ref{sec:analysis}.

\subsection{Maximum likelihood analysis of the sample}
\label{sec:mla}

\begin{figure*}
\centering\includegraphics[width=0.75\hsize]{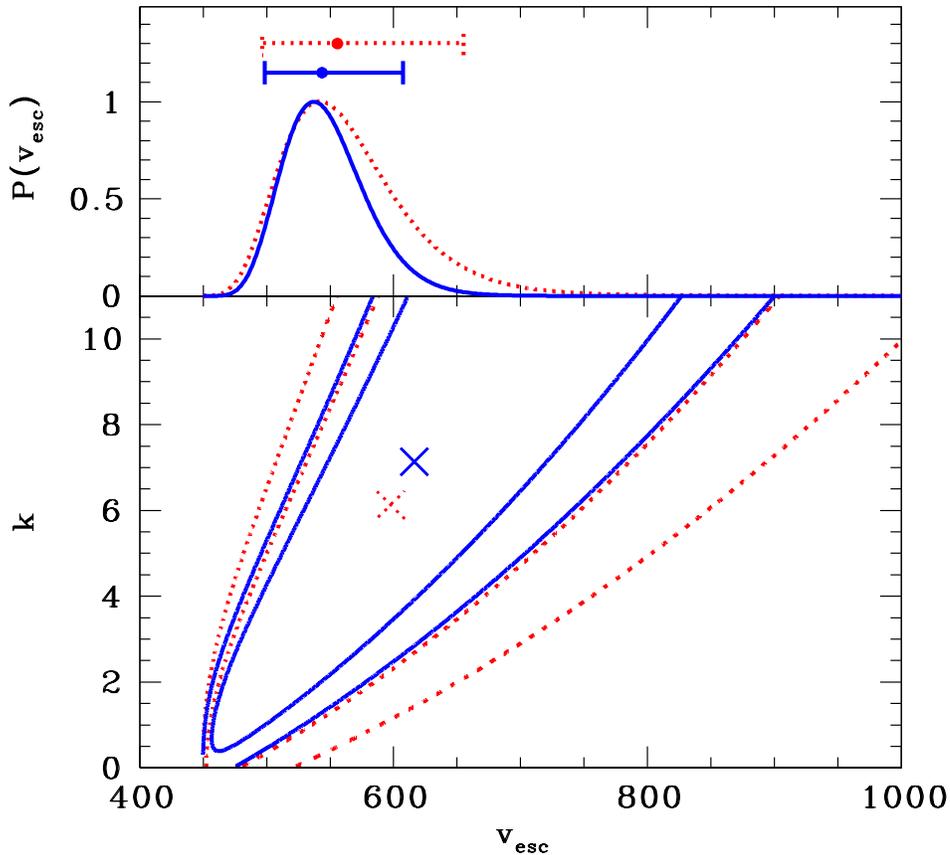}
\caption{The lower panel shows the 2-dimension likelihood contours
that can be placed on the local escape velocity ($\ve$) and the shape
of the velocity distribution ($k$; see Section \ref{sec:analysis})
from our combined high-velocity sample. The cross corresponds to the
peak likelihood, while the contours denote 10 and 1 per cent of this
peak likelihood value. The upper panel shows the likelihood
distribution for $\ve$ obtained by assuming a uniform prior on
$k\in[2.7,4.7]$ (see Section \ref{sec:abadi}); the corresponding error
bar shows the 90 per cent confidence interval. The dotted quantities
show the results from a sample containing only the high-velocity RAVE
stars, i.e. a smaller sample of 16 stars.}
\label{fig:contours}
\end{figure*}

Evaluating equation (\ref{eq:lhood}) results in 2-dimensional
likelihood contours for $k$ and $\ve$, shown in the lower panel of
Fig. \ref{fig:contours}. The peak of the likelihood lies at
$\ve=616\kms$ and $k=7.1$. Although this value of $k$
is slightly greater than that predicted from our cosmological simulations in
Section \ref{sec:abadi}, we find that our peak likelihood is
relatively broad. In addition, as was discussed in Section
\ref{sec:quantify_contam}, although our predicted level of thick-disc
contamination should have little effect on our $\ve$ constraints,
any contamination will shift the peak likelihood to higher values of $k$.
In Fig. \ref{fig:data} we show the probability
distribution for the radial velocities (given by equation
\ref{eq:vrdist}) corresponding to this peak likelihood.

As can be seen from Fig. \ref{fig:contours}, given the size of our
sample of high radial velocity stars it is not possible to
constrain simultaneously both $\ve$ and $k$ using this method.

From this figure we can see that there is a degeneracy between $k$ and
$\ve$. This can be understood as follows: 
to retain the same shape of the velocity distribution over the
observed range of velocities, an increase in $\ve$ must be accompanied
by an increase in the steepness of the slope of the distribution
(i.e. $k$).
Given this degeneracy, we need to apply a suitable prior on $k$ to
obtain constraints on $\ve$. The results from Section
\ref{sec:abadi} indicate that a suitable choice is to adopt a uniform
prior in the range $k\in[2.7,4.7]$. We have done this, and the results
are shown in the top panel of Fig. \ref{fig:contours}. The resulting
90 per cent confidence interval is, \beq
\label{eq:vecons}
498\kms < \ve < 608 \kms, \eeq with a median likelihood of
$\ve=544\kms$.
Note that despite the degeneracy between $k$ and $\ve$, it is
immediately evident that negative values of $k$ are strongly disfavoured.

It is important to check whether any kinematical bias is introduced
by combining data from separate surveys. To do
this we repeat the above analysis using only the 16 RAVE stars from
Section \ref{sec:ravefin}. This is shown by the dotted quantities in
Figure \ref{fig:contours}. Reassuringly there is no noticeable offset
between the two sets of contours; the only difference is a general
broadening of the contours. When we apply the prior $k\in[2.7,4.7]$
we find that the 90 per cent confidence interval becomes $496 < \ve
< 655 \kms$, with a median likelihood of $\ve=556\kms$.

\subsection{Bootstrap analysis}
\label{sec:br}

To further assess the likelihood constraints presented in the previous
section, we also apply the bootstrap technique (see Section
\ref{sec:boot}) to our data.

We apply the bootstrap approach to the combined dataset of 33 stars,
but unlike Section \ref{sec:mla} we apply the LT90 approximation
(equation \ref{eq:vrdistlt}) when calculating the maximum
likelihood. The bootstrap computed the values of $\ve$ and $k$ that
maximized equation (\ref{eq:lhood}) using 5000 resamples of the
original RAVE sample. Table \ref{tab:bootdata} shows the resulting
values of $\ve$ and $k$ for the two chosen priors (see Appendix
\ref{appendix}).

\begin{figure}
\plotone{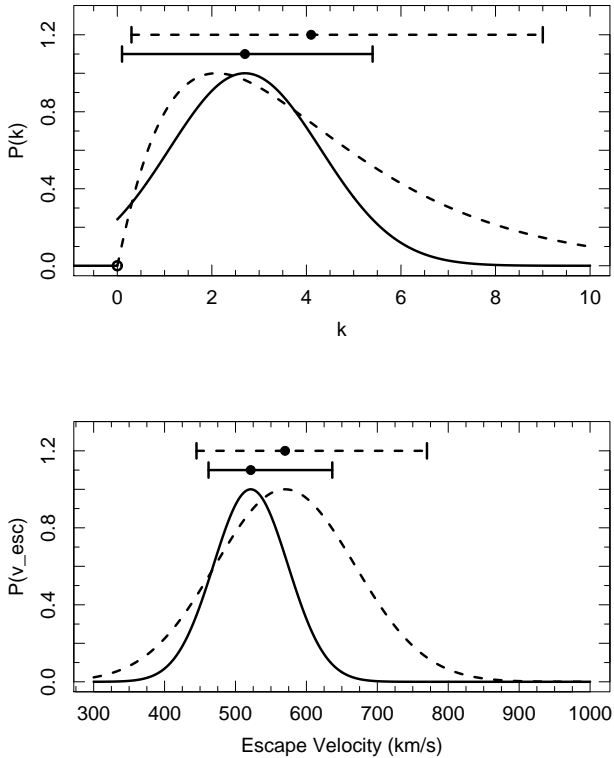}
\caption{
The bootstrap distributions for $k$ (top panel) and $\ve$
(lower panel). The dashed and solid curves correspond to the
bootstrap distributions calculated when solving equation (\ref{eq:lhood}) with
Priors 1 and 2, respectively.  The corresponding error bars show the
90 per cent confidence intervals for the parameters for each prior.
Note that for both parameters, Prior 2 gives the smallest confidence
interval. Furthermore, note that the bootstrap distribution for $k$
follows a chi-squared distribution.}

\label{fig:bootconf}
\end{figure}

When we compare the bootstrap interval with the likelihood interval
obtained in Section \ref{sec:mla}, we find that the interval is
shifted towards smaller $\ve$. This is consistent with what one would
expect for the bootstrap method, since (unlike the method described in
Section \ref{sec:mla}) the process of bootstrapping can result in
values of $\ve$ that are smaller than the highest velocity star in the
sample. This is a consequence of the fact that the bootstrap approach
accounts for possible unreliable or inconsistent data.  However, it is
also important to note that the bootstrap mean values of $k$ and $\ve$
found with both priors are identical, within standard errors, to those
found in the previous section using the non-bootstrap technique with
the LT90 prior. Figure \ref{fig:bootconf} shows the bootstrap
distributions and corresponding confidence intervals calculated for $k$
and $\ve$ when each prior is applied to equation (\ref{eq:lhood}).
The dashed curves correspond to Prior 1, while the solid curves
correspond to Prior 2.  The confidence intervals obtained using Prior
2 are clearly smaller than that from Prior 1, owing to the fact that
Prior 2 contains more information about our expectations of $k$.

As a result of our analyses with a simulated dataset (see Appendix
\ref{appendix}), we adopted the confidence regions from Prior 2,
obtaining the bootstrap 90 per cent confidence intervals $462~\kms <
\ve < 640~\kms$ and $0.1 < k < 5.4$. 

\begin{table*}
\begin{tabular}{cllccc}
\hline Label & Prior Form & Initial MLE & Bootstrap & SE &
90\%-Conf.\\
 & & Values & Mean Values & & \\
 \hline
1 & $P_{LT}(\ve,\,k) \propto 1/\ve$ & $\ve=588.7\kms$ & 569.9 & 98.3 &
(445.0 769.4) \\
 & & $k=4.2$ & 4.1 & 3.0 & (0.3,9.0) \\
\\
2 & $P_J(\ve,\,k) \propto \frac{\sqrt{k}}{(\ve-\vmin)\sqrt{k+2}}$
& $\ve=535.2\kms$ & 521.5 & 52.5 & (462.0,640.0) \\
 & & $k=2.9$ & 2.7 & 1.6 & (0.1,5.4)\\
\\
\hline
\end{tabular}
\caption{Bootstrap results for combined data with $\vmin=300\kms$. The
 third column gives the maximum likelihood values of $\ve$ and $k$
 for the original (non-bootstrapped) sample, while the fourth column
 gives the mean of the bootstrap distribution of maximum likelihood
 values (Note that the means quoted here for Prior 1 differ from those
 given for the maximum likelihood method in Section \ref{sec:mla}
 since the former employs the LT90 approximation, i.e. equation
 \ref{eq:vrdistlt}).
 The SE column gives the standard errors from the standard 
 deviation of the bootstrap distribution. The last column gives the 90
 per cent confidence intervals for $\ve$ and $k$.  The $\ve$
 confidence intervals are computed using the bootstrap normal-distribution
 approximation, while the $k$ intervals are calculated from a
 chi-squared distribution with the number of degrees of freedom equal
 to the bootstrap mean value.}
\label{tab:bootdata}
\end{table*}

\section{Discussion}
\label{sec:disc}

\subsection{The nature of the fastest RAVE stars}
\label{sec:rpmd}

One might wonder about the nature of the fastest stars in the RAVE
catalogue. Although we do not currently possess any additional
information from the RAVE spectra, such as metallicities, we do have
estimates for each star's proper motion from various sources
(Steinmetz et al. 2006).
When combined with accurate photometry (in our case, $J$-
and $K$-band magnitudes from 2MASS [Skrutskie et al. 1997]), proper
motions can be used to place the stars on a reduced proper motion
(RPM) diagram. A RPM diagram can be used to differentiate types of
stars such as dwarfs and giants of different stellar populations
because it is closely related to a standard colour-magnitude diagram,
modified by kinematics. The $J$-band RPM is given by \beq J_{\rm RPM}
\equiv J + 5 {\rm log} \mu + 5= M_J + 5 {\rm log} 
\frac{v_{tangential}}{47.4 \kms} + 5, \eeq where $\mu$ is the proper
motion in arcsec yr$^{-1}$ and $v_{tangential}$ is the tangential
velocity corresponding to this proper motion. Thus one obtains a
quantity that is related to the absolute magnitude, but with stellar
populations of different mean motions offset from one another, with
scatter induced by the velocity dispersion.

\begin{figure}
\centering\includegraphics[keepaspectratio=true,width=2.5in,angle=90]{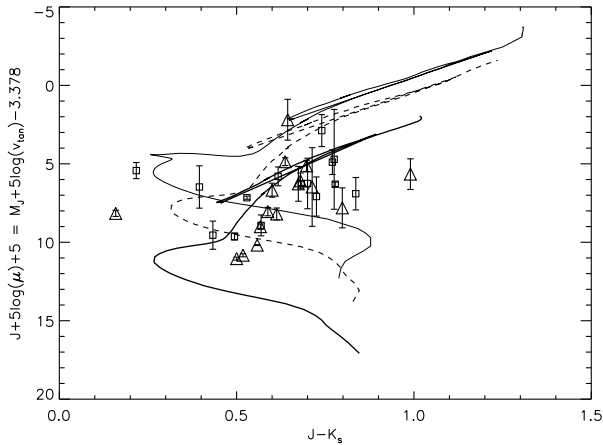}
\caption{J-Band reduced proper motion diagram of RAVE stars selected
to have Galactocentric radial velocity greater than $250\kms$. The
triangle symbols represent stars with Galactocentric radial velocity
greater than $300\kms$. Note that the errors in the photometry are
smaller than the symbols. Isochrones from Bonatto et al. (2004) are
overplotted, with different tangential velocities adopted, to
represent the three main stellar components of the galaxy. The thin
solid line is appropriate for the thin disk ($Z=0.019$, age$=2.5$ Gyr,
$v_{tangential}=20\kms$), the dashed line describes the thick disk
($Z=0.004$, age$=10$ Gyr, $v_{tangential}=42\kms$), and the thick
solid line describes the halo ($Z=0.001$, age$=10$ Gyr,
$v_{tangential}=200\kms$). The majority of the stars have positions in
this diagram consistent with being halo giants. However, the star with
$J-K$ corresponding to $\sim 0.15$ is most likely a BHB star.}
\label{fig:rpmd}
\end{figure}

We show the RPM for our fastest RAVE stars in Fig \ref{fig:rpmd}.
Three Bonatto et al. (2004) isochrones, adjusted for different mean
$v_{tangential}$, are also plotted. The three isochrones have ages and
metallicities corresponding to different assumptions regarding the
kinematics: namely thin disk, thick-disk and halo populations. The
thin solid isochrone has been plotted assuming thin disk kinematics,
with tangential velocity $v_{tangential}=20\kms$, $Z=0.019$, and
age$=2.5$ Gyr. The dashed isochrone represents the thick-disk
component, $v_{tangential}=42\kms$, $Z=0.004$, age$=10$ Gyr. Finally,
the thick solid isochrone depicts the halo with
$v_{tangential}=200\kms$, $Z=0.001$, and age$=10$ Gyr.  In a
probabilistic assignment of any one star to a given population, one
would take account of the expected $\sigma \sim 150\kms$ spread around
the halo isochrone.

The RAVE high-velocity stars clearly scatter around the red giant
branch of the halo isochrone; this association is even stronger for
the highest-velocity stars, those with $\vr > 300\kms$, denoted by the
triangle symbols. This is reassuring because it indicates that our
sample should not suffer from significant contamination from the thick
disc, thus justifying our choice of $\vmin$ in Section \ref{sec:vmin}.
Furthermore, using the echelle data from APO (section
\ref{sec:validation}), we were able to calculate gravities and
temperatures for a few of the high-velocity stars. These gravities
were low, confirming giants. The bluest star appears from its spectrum
to be a halo blue horizontal branch (BHB) star.\footnote{The BHB star has
identifier, C1100242-024226.}

If these stars are indeed mostly halo giants, then they are not in the
immediate solar neighbourhood. This could have implications for our
determination of $\ve$ and so we investigate this issue in the
following section.

It is also worth noting that our sample contains three relatively high
proper-motion stars with $\mu>50$ mas yr$^{-1}$ (C1022369-140345,
C1250398-030748 and T8395\_01513\_1). These all appear to be halo
dwarfs or relatively nearby halo giants.
By calculating distances from either our follow-up spectra or from the
crude distance calibration of Scholz, Meusinger \& Jahrei{\ss} (2005)
we find that their total 3D velocities are all well within our value
of $\ve$.

\subsection{The effect of sample volume on the escape velocity}
\label{sec:volume}

In the Section \ref{sec:abadi} we used a sample of stellar particles
from cosmological simulations to analyse our ability to recover the
escape velocity. In that section our solar neighbourhood volumes were
constructed from spheres of radius 2 kpc centred at 8.5 kpc (and also
equivalent spheres of radius 1 kpc and 3 kpc centred at 4.25 kpc and
12.75 kpc, respectively, thus preserving the ratio of sample radius to
distance from the Galactic centre). However, the catalogue of radial
velocity measurements used to try to constrain the Milky Way escape
velocity could contain stars at significantly larger distances. The
magnitude limit of the RAVE survey is $I=12$ mag, which implies that
very bright giants may be a significant distance away; for example, a
K2 giant with $M_I\approx-0.9$ mag could be at a distance of around 4
kpc. Indeed, Section \ref{sec:rpmd} suggested that some of these
fastest RAVE stars may well be halo giants.

We test whether our results are sensitive to the sample volume by
analysing the KIA3 and KIA4 simulations introduced in Section
\ref{sec:abadi} (we restrict ourselves to these two galaxies as they
have the largest number of stellar particles). We generate samples of
stellar particles by taking spheres of varying radius centred on 8.5
kpc. From these samples we can estimate the escape velocity by
evaluating equation (\ref{eq:lhood}), adopting the prior determined in
Section \ref{sec:abadi} (i.e. $k \in [2.7,4.7]$). This analysis shows
how the recovered escape velocity varies as a function of the radius
of the `solar neighbourhood' spheres. We find that for larger spheres
the escape velocity is slightly over-estimated: the error in $\ve$ is
$\la 15$ per cent for all radii up to 5 kpc, with an error of $\la 10$
per cent for radii of less than 4 kpc. This over-estimation is
probably due to the fact that larger volumes include stars that probe
regions of higher $\ve$ and hence result in larger estimates for
$\ve$. It should also be noted that in reality this is an upper-limit
since only a small proportion of our sample of observed stars will
probe such large distances, whereas these simulated samples do not
incorporate any such distance effects (i.e. the probability of
selecting a simulated star is independent of its distance).

\subsection{Unbound stars}
\label{sec:unbound}

One obvious deficiency of the analysis presented in Section
\ref{sec:analysis} is that it requires all stars in our sample to be
bound to the Galaxy. It is clear that the presence of any unbound
stars would invalidate our results, although the bootstrap technique
should reduce our sensitivity to any spurious datapoints. There are
currently seven known `hyper-velocity', probably unbound, stars in our
Galaxy (Edelmann et al. 2005; Hirsch et al. 2005; Brown et
al. 2006). The Galactic standard of rest frame radial
velocities computed for these stars range from $\sim 550\kms$ to $\sim
720\kms$. The first few were found serendipitously while the most
recent were found by a targeted survey of faint early-type stars
(Brown et al.~2006); indeed all of the known hyper-velocity
stars are blue and therefore almost certainly young (Kollmeier \&
Gould 2007).

It is suggested that such high velocities may originate from
encounters with the massive black hole at the Galactic
centre. However, Yu \& Tremaine (2003) found that the
numbers of hyper-velocity stars produced would be about $10^{-5} \rm
yr^{-1}$ (see also Perets, Hopman \& Alexander 2006). That would give
about $10^5$ stars in total, assuming a long-lived black hole, which
suggests that hyper-velocity stars are very rare. Furthermore, they
found that the number of lower velocity `high-velocity' stars expected
is even smaller. Although such objects should be rare, and hence the
chance of one appearing in our catalogue is small, this can
nevertheless not be ruled out entirely. There is one simple test one
can apply to investigate this issue; if one can obtain the full 3
dimensional velocities for the fastest stars, then it can be
ascertained whether their direction of motion is consistent with that
of an object ejected from the centre of the Galaxy. However, this
method could fail if the unbound star(s) originate from different
circumstances, such as stars which are bound to the local group but
unbound with respect to our Galaxy.

High-velocity stars can also originate from stellar binary systems in
which one of the stars undergoes a supernova and `kicks' out the
companion star.  This would have larger effects on main sequence stars
as they can acquire an extra velocity of $\sim100\kms$.  However,
since our sample contains mostly halo giants (see Section
\ref{sec:rpmd}) this should be a small effect.

One final point to note is that given the consistency between our
constraints on $\ve$ with other independent methods for estimating the
mass of the Galaxy, the presence of any unbound stars in our sample
seems to be an unlikely scenario.

\subsection{The mass and extent of the Galactic halo}
\label{sec:halomass}

The escape velocity can be used as a powerful tool to probe the mass
distribution in the Galaxy. Our constraints on $\ve$ allow us to
constrain the halo potential at the Solar radius,
\beq
\Phi_{\rm total}(r_\odot)=-\ve^2/2.
\eeq
Halo masses derived from our constraints on $\ve$ rely on the
assumption that our fastest stars probe velocities all the way to the
local escape velocity, i.e. there is no truncation in the stellar
velocity distribution. If this assumption is invalid then the mass
constraints quoted in this section would be lower-limits to the true
mass. However, as has been shown for the simulated galaxies in Section
\ref{sec:abadi}, the level of truncation in the stellar component cannot
be more than 10 per cent. When the likelihood analysis (equation
\ref{eq:lhood}) is applied to these simulations we do not find any
systematic underestimation in the recovered $\ve$, which gives us
confidence that such weak levels of truncation (if present) will not
pose significant problems for our analysis.

extend up to 
When we determine the escape velocity for the s
the maximum likelihood estimator for 

In this section we will use $\ve$ to constrain the total mass of the
halo. However, before undertaking any detailed calculations, one
obtain a qualitative understanding of the situation from the simple
relation (see, for example, equation 2-22 of Binney \& Tremaine
[1987]),
\beq
\ve^2=2 \, \vc^2 + 8 \pi G \int_{r_\odot}^\infty \rho(r) \, r \,
\mathrm{d}r,
\label{eq:mass_int}
\eeq
where $\vc$ is the circular velocity at the Solar radius. The
fact that our measured lower-limit of $\ve > 498\,\kms$ is 
significantly greater than $\sqrt{2} \vc \sim 311 \kms$ shows that the
second term in equation (\ref{eq:mass_int}) 
cannot be small, i.e. there must be a significant amount of mass
exterior to the Solar circle. This simple argument convincingly
demonstrates the presence of a dark halo in the Galaxy.

Another straightforward calculation that can be made is the extent of
an isothermal halo that has a constant circular velocity out to a
truncation radius $r_{\rm cut}$. For this elementary model we find that,
\beq
r_{\rm cut} = r_\odot \;{\rm exp} \left[\frac{\ve^2}{2\vc^2}-1\right].
\eeq
Our measured value of $\ve$ ($\ve\approx544\kms$) corresponds to a
truncation radius of $r_{\rm cut}\approx 58$ kpc.

The above simple arguments demonstrate the existence of a dark halo,
but we now undertake a more detailed calculation in order to constrain
the total mass of the halo. To do this we first need a model for the
baryonic contribution to the total potential from the bulge and disc
of the Galaxy. We adopt a spherical Hernquist (1990) bulge with mass
$1.5\times10^{10}\sm$ and scale radius 0.6 kpc, and a Miyamoto \& Nagai
(1975) disc with mass $5\times10^{10}\sm$, scalelength 4 kpc and
scaleheight 0.3 kpc. The halo potential is then simply,
\beq
\Phi_{\rm halo}=\Phi_{\rm total}-\Phi_{\rm bulge}-\Phi_{\rm disc}.
\eeq

When $\Phi_{\rm halo}$ is combined with a halo profile it is possible
to probe the total halo mass. One popular profile is the NFW halo
(Navarro, Frenk \& White 1996). It can be shown that the radial potential for
an NFW density profile can be expressed as,
\beq
\Phi_{\rm NFW}(r)=-\frac{4\pi G \rho_s r_{\rm vir}^3}{c^3 r} {\rm
ln} (1+\frac{c r}{r_{\rm vir}}),
\eeq
where $c$ is a concentration parameter ($c$) equal to the ratio of the
virial radius to the scale radius, and $\rho_{\rm s}$ is a
characteristic density given by,
\beq
\rho_s=\frac{\rho_{\rm cr} \Omega_0 \delta_{\rm th}}{3}
\frac{c^3}{{\rm ln}(1+c)-c/(1+c)},
\eeq
where $\rho_{\rm cr}=3H^2/8\pi G$ is the critical density of the
universe, $\Omega_0$ is the contribution of matter to the critical
density and $\delta_{\rm th}$ is the critical overdensity at
virilisation. The virial mass can then be determined from the virial
radius,
\beq
M_{\rm vir}=\frac{4\pi}{3} \rho_{\rm cr} \Omega_0 \delta_{\rm th}
r_{\rm vir}^3.
\eeq 

If we take the solar radius to be 7.5 kpc, $\Omega_0=0.3$,
$\delta_{\rm th}=340$, $H=65\kms\,{\rm Mpc}^{-1}$, and enforce the
circular velocity at the solar radius to be $\vc=220\kms$, our 90 per cent
confidence constraint on the escape velocity (equation
\ref{eq:vecons}) allow us to obtain the following constraints on the
virial radius, mass and concentration parameter for the Milky Way,
\begin{displaymath}
M_{\rm vir}=0.85^{+0.55}_{-0.29}\times10^{12}M_\odot,
\end{displaymath}
\begin{displaymath}
r_{\rm vir}=257^{+47}_{-33}\,{\rm kpc},
\end{displaymath}
\begin{displaymath}
c=24.3^{+6.5}_{-5.1}.
\end{displaymath}
It is worth noting that these constraints have little dependence on
the adopted value for the solar radius; taking a value of 8 kpc will
change the limits on $M_{\rm vir}$ and $r_{\rm vir}$ by less than 1 per cent.

One common variation of this classical NFW model is to account for the
adiabatic contraction of the dark matter halo due to the presence of
the baryons in the disc and bulge (Mo, Mao \& White 1998). In this
instance we simplify the calculation by adopting an exponential disc,
retaining the same mass and scalelength as above. The resulting
constraints on the virial mass, radius and concentration are,
\begin{displaymath}
M_{\rm vir}=1.42^{+1.14}_{-0.54}\times10^{12}M_\odot,
\end{displaymath}
\begin{displaymath}
r_{\rm vir}=305^{+66}_{-45}\,{\rm kpc},
\end{displaymath}
\begin{displaymath}
c=7.7^{+2.8}_{-2.0}.
\end{displaymath}

Both of these models predict low values for the virial velocity of the
halo. The contracted NFW model has $\vv=142^{+31}_{-21}\kms$, while the
uncontracted one has $\vv=124^{+20}_{-14}\kms$, i.e. around half $\vc$.
These low values of $\vv$ pose problems for semi-analytic models of
galaxy formation that require the peak circular velocity of the
disc to be $\sim \vv$ in order to simultaneously explain the
normalisation of the Tully-Fisher relation and the galaxy luminosity
function in a $\Lambda$CDM universe (e.g. Cole et al. 2000).
More recent semi-analytic models (Croton et al. 2006a,
2006b) have been able to relax this requirement slightly, so that
the peak circular velocity of the disc is only required to be
comparable to the maximum circular velocity of the dark matter
halo. However, it can be seen that our models are only marginally
consistent with this requirement, e.g. for our contracted model, the
peak circular velocity of the halo is only 167 $\kms$ while the peak
circular velocity of the disc is 220 $\kms$.

It is possible to work out similar constraints for other halo
models. For example, one can take the model of Wilkinson \& Evans
(1999), which has a flat rotation curve in the inner regions and a
sharp fall off in density beyond an outer edge (hereafter referred to
as the WE model). For this halo profile the total halo mass is given by,
\beq
M_{\rm WE}=\frac{\ve^2}{2G}
a_{\rm WE} \left( {\rm ln} \left[ \frac{\sqrt{r^2+a_{\rm WE}^2} +
a_{\rm WE}}{r_\odot} \right] \right) ^{-1},
\eeq
where $a_{\rm WE}$ is the scalelength. For this model we revert to our
original Miyamoto \& Nagai (1975) disc, as used for the uncontracted
NFW profile. By again enforcing the circular velocity to be
$220\kms$ we can use this model to obtain the following 90 per cent
confidence constraints on the total mass and scalelength,
\begin{displaymath}
M_{\rm WE}=1.89^{+76.46}_{-1.13}\times10^{12}M_\odot.
\end{displaymath}
\begin{displaymath}
a_{\rm WE}=314^{+986}_{-188}\,{\rm kpc},
\end{displaymath}

It is clear that our constraints on the halo mass are consistent with
previous estimates; within the last ten years estimates tend to lie in
the range $1-2\times 10^{12}\sm$ (see Section \ref{sec:intro}), which
is fully consistent with our findings. It is also important to note
that our constraints are relatively model independent, with only small
differences between the mass estimates for our three halo profiles.

It is also useful to quote the total mass within a certain radius of
the Galactic centre. For example, our three models predict the
following constraints on the mass contained
within 50 kpc:
$4.04^{+1.10}_{-0.76}\times10^{11}M_\odot$,
$3.87^{+0.64}_{-0.56}\times10^{11}M_\odot$,
$3.58^{+0.04}_{-0.17}\times10^{11}M_\odot$,
for the uncontracted NFW, contracted NFW, and WE model,
respectively. These masses are relatively well constrained mainly due to the
fact that our models must adhere to the $\vc=220\kms$ constraint.
Wilkinson \& Evans (1999) obtained a value of
$5.4^{+0.2}_{-3.6}\times10^{11}M_\odot$, which is consistent with our
models.

Given these halo models, we can estimate the maximum Galactic radius that
our fastest star will reach. From Table \ref{tab:rave} we see that the
fastest star has a rest-frame radial velocity  of $-449\kms$. We
assume that the star's kinetic energy is dominated by its radial
component, so
\beq
\label{eq:esun}
E(r_\odot)=(\vr^2-\ve^2)/2.
\eeq
At apocentre the kinetic energy is solely due to the angular momentum
$L$, which we assume to be conserved. Assuming that the distance to
the Sun is small, we have
\beq
L = \left(
\begin{array}{ccc}
7.5~{\rm kpc}\\ 0\\ 0\\
\end{array}
\right) \times \left(
\begin{array}{ccc}
- \vr {\rm cos}~b~{\rm cos}~l \\ \vr {\rm cos}~b~{\rm sin}~l \\ \vr
{\rm sin}~b\\
\end{array}
\right),
\eeq
The apocentre distance $\rapo$ can now be found from
\beq
\label{eq:eapo}
E(\rapo)=\frac{L^2}{2\rapo^2}+\Phi(\rapo).
\eeq
Equations (\ref{eq:esun}) and (\ref{eq:eapo}) can be solved using any
of the models for the potential described above. For our constraints
on $\ve$ determined in Section \ref{sec:mla},
we find that $\rapo = 102^{+51}_{-22}$ kpc for our contracted NFW halo
model. Since this calculation assumes that there is no tangential
velocity, it is actually a lower-limit for $\rapo$; for example, if
the star has a tangential velocity of $100\kms$ then $\rapo$ increases
to $\sim 122^{+81}_{-31}$ kpc.
Although we do not know the distance to our fastest radial velocity
star, by adopting a distance of 1 kpc we find that the observed proper
motion ($17.2\pm9.4$ mas yr$^{-1}$; Hambly et al. 2001) corresponds to
a tangential velocity of $82\pm46\kms$.

The question of whether there is a truncation in the stellar halo is
currently open to debate (e.g. Ivezi\'{c} et al. 2000; Dehnen et
al. 2006). Therefore our large value for $\rapo$ is important since it
could disfavour Galactic models with an early truncation.

\subsection{Uncertainties and the next steps}
\label{sec:future}

As discussed in the Section \ref{sec:analysis}, our results are based
on the assumption that the velocity distribution near the escape
velocity can be approximated by a power-law $f(v) \propto
(\ve-|{\bf v}|)^k$. In a hierarchical universe, this model is
unlikely to be valid because the motions of the fastest moving stars
are expected to be strongly clumped (Helmi et al. 2002). Since the velocity
distribution would be dominated by very few streams, it would be the
sum of a few delta functions centered on the mean velocities of those
streams.  It may even be possible that each stream originates in the
same object, in which case, only one orbit is probed. This invalidates
the use of statistical arguments to derive the mass of the Galaxy.
Such a high-degree of lumpiness only is important when the fastest 1
per cent of the stars are considered, that is, those with velocities
$\vr \ga 3 \sigma \sim 450 \kms$.

Notice that there are no stars in our sample with such large radial
velocities.  This is the reason why substructure appears not be an
issue in our case (for example, the results of the bootstrap analysis
obtained by re-sampling are consistent with the maximum likelihood
performed using the actual dataset). However, a new approach will be
necessary in the future. By the time the RAVE survey concludes, it is
intended that the catalogue will contain up to one million stars. If
we assume that the rate of detecting high velocity stars is unchanged
for the rest of the survey, this would result in a final sample of
$\sim200$ stars with radial velocities greater than $300\kms$. Such a
sample will reveal a large amount of substructure.  This should also
be evident also when the RAVE metallicities are added.  Gravity
determinations, plus the metallicities, will allow robust distance
estimates and hence the use of tangential velocities. Full phase-space
information should enable one to perform more sophisticated models to
determine the mass of our Galaxy.

All of this highlights the importance of understanding the expected shape
of the velocity distribution near the tails (Section \ref{sec:abadi}).
The limited number of high-resolution cosmological simulations of
galaxies like the Milky Way is also a source of uncertainty in our
estimates of the escape velocity. Both higher resolution (a
significantly larger number of stellar particles) and a larger set of
simulations are necessary.

\section{Conclusion}
\label{sec:conclusion}

In this work we have reported new constraints on the local escape
speed of our Galaxy. We argued that the choice of prior on $k$ may
have been incorrectly estimated in previous work and deduced a
different range for this prior (see Section \ref{sec:abadi}). We then
applied this to a catalogue of radial velocities from the RAVE
collaboration, augmented with additional stars from the existing
survey of Beers et al. (2000). Our results provide a 90 per cent
confidence interval of $498\kms < \ve < 608 \kms$, with a median
likelihood of $\ve=544\kms$. 
We also applied a bootstrap technique, which allowed us to
reduce dependencies on extreme velocities that may be unbound or from
stars in binary systems, while additionally allowing us to perform
confidence analysis on the distribution of maximum likelihood values,
$\ve$ and $k$, simultaneously.  This resulted in a 90 per cent confidence
interval of $462~\kms < \ve < 640\kms$ and $0.1 < k < 5.4$.

The fact that $\ve^2$ is significantly greater than $2\vc^2$ implies
that there must be a significant amount of mass exterior to the Solar
circle. Furthermore, from our constraints on $\ve$ we can infer model
dependent estimates for the mass of the halo (see Section
\ref{sec:halomass}); for example, an adiabatically contracted NFW halo
profile predicts that
$M_{\rm vir}=1.42^{+1.14}_{-0.54}\times10^{12}M_\odot$,
$r_{\rm vir}=305^{+66}_{-45}\,{\rm kpc}$, and
$c=7.7^{+2.8}_{-2.0}$.
Similar results for the halo mass were found for both an uncontracted NFW
halo and a Wilkinson \& Evans (1999) halo. Although our results are
model dependent, we find that our three models are in good agreement.
The mass within 50 kpc is found to be
$3.6-4.0\times10^{11}M_\odot$. It is interesting to note that
our models predict low values for the virial velocity, for example
$\vv=142^{+31}_{-21}\kms$ for the contracted NFW model.

By the time the RAVE survey reaches completion (currently predicted to
be $~$2010) it will provide an unparallelled database of stellar
kinematics, thus allowing dramatic progress in this field.

\section*{acknowledgments}

The authors wish to thank M. Abadi for advice and support regarding
the cosmological simulations utilised in Section \ref{sec:assess},
A. Villalobos for assistance with Section \ref{sec:halomass} and
P. Maiste and D. Naiman for their assistance regarding the bootstrap
analysis discussion.
The authors are indebted to S. White and the anonymous referee for
comments that greatly improved the clarity of the paper.
MCS and AH acknowledges financial support from the Netherlands
Organisation for Scientific Research (NWO). GR \& RFGW acknowledge
financial support from the US National Science Foundation through
grant AST-0508996.  This work benefited from the support of the
European Community's Sixth Framework Marie Curie Research Training
Network Programme, Contract No. MRTN-CT-2004-505183 ``ANGLES".

Funding for RAVE has been provided by the Anglo-Australian
Observatory, by the Astrophysical Institute Potsdam, by the Australian
Research Council, by the German Research Foundation, by the National
Institute for Astrophysics at Padova, by The Johns Hopkins University,
by the Netherlands Research School for Astronomy, by the Natural
Sciences and Engineering Research Council of Canada, by the Slovenian
Research Agency, by the Swiss National Science Foundation, by the
National Science Foundation of the USA, by the Netherlands
Organisation for Scientific Research, by the Particle Physics and
Astronomy Research Council of the UK, by Opticon, by Strasbourg
Observatory, and by the Universities of Basel, Cambridge, and
Groningen.

The RAVE web site is at www.rave-survey.org.

\appendix
\section{The prior probability distributions for the bootstrap method}
\label{appendix}

\subsection{The choice of prior}
\label{sec:prior_choice}

\subsubsection{Jeffreys' rules}

It can be quite difficult to choose \textit{a priori\/} probabilities
for $\ve$ and $k$ especially when not much is known about the
quantities. Note that the posterior distribution (equation
\ref{eq:lhood}), with uniform prior distributions (i.e.,
$P(\ve)=P(k)=1$), is equivalent to the normalised likelihood equation.
Furthermore, even if the prior probabilities are non-uniform, the
maximization of equations (\ref{eq:lik}) and (\ref{eq:lhood}) are
asymptotically equivalent. This is due to the fact that as $n$ goes to
infinity, the product in equation (\ref{eq:lhood}) will dominate.
However, for small $n$, there may be large differences between the
parameter values maximizing $l(\ve,k)$ and those maximizing the
posterior distribution. Therefore it is important, for small $n$, to
choose `good' reference priors for $\ve$ and $k$.

One method that can be used to attain reference priors is that defined
by Sir Harold Jeffreys (Jeffreys 1961). First define the log of the
normalised likelihood equation (equation \ref{eq:lik}) as
$L(\ve,k)$. Next, we define a $2\times2$ information matrix, known as
the Fisher Information matrix:
\begin{equation}
I(\bm{\theta} | \mathbf{\vr})_{i,j} = - E
\left[\frac{\partial^2 L(\ve,k)}{\partial \theta_i \,\partial \theta_j} \right]
\label{eq:fishinf}
\end{equation}
where $\bm{\theta}=(\ve,\,k)$ and the right hand side is the negative of
the expectation value of the second derivative. Note that if ${\bmath
\vr}$ is a vector of $n$ independent observations, by linearity of
expectation the information is $nI(\bm{\theta} | \vr)_{i,j}$. However, we
do not include this factor $n$ since it is a constant and will not affect
the maximum likelihood. Using this information matrix, Jeffreys argues
that a good reference prior can be estimated as,
\begin{equation}
P_J(\theta) \propto \sqrt{\left| I({\bm \theta} |\vr) \right|},
\label{eq:pri}
\end{equation}
where $\left| I({\bm \theta} |\vr) \right|$ is the determinant of the
information matrix.  This equation is a useful reference prior because
it does not require the selection of any specific parametrization.
However, it is often argued that this form of the prior is improved by
considering that the parameters are independent of each other (Lee
2004). We follow this approach, which requires us to neglect the
off-diagonal terms in the determinant.  By applying this analysis to
the LT90 formalism (equation \ref{eq:vrdistlt}) we obtain,
\beq
P_J(\ve,\,k) \propto \frac{1}{(\ve-\vmin)\sqrt{k(k+2)}}.
\label{eq:pri_J}
\eeq

LT90 apply another of Jeffreys' Rules that says if a variable varies
continuously from 0 to infinity, one must adopt an {\em a priori}
probability distribution equal to the reciprocal of the variable
(Kendall \& Stuart 1977). LT90 adopt this form of the prior
probability distribution for the escape velocity, $P_{LT}(\ve) \propto
1/\ve$. However, note that $\ve$ does not necessarily vary between
(0,$\infty$). We have applied a minimum cut-off velocity, as defined
in Section \ref{sec:vmin}. Therefore, an arguably more appropriate
approach would be to say that the escape velocity varies continuously
in $(v_{min}$,$\infty)$, i.e. $P_{J^\prime}(\ve) \propto
1/(\ve-\vmin)$.  This approach can also be applied to the parameter
$k$. From equation (\ref{eq:vrdistlt}) we can see that for the LT90
formalism the variable $k$ varies between $(-1,\infty)$ and so one
could also adopt the prior $P_{J^\prime}(k) \propto 1/(k+1)$.

\subsubsection{{\em A priori} probability}
The question remains, which prior probabilities should be used? In an
ideal world, the samples should be large enough so that
maximization of equation (\ref{eq:lhood}) would be the same for any
chosen prior. However, we cannot rely on this fact. It is very
important to think of Jeffreys' Rules as only guidelines for choosing
prior distributions, especially when dealing with more than one
parameter. Therefore we have chosen three priors for our analysis. The
first is that used by LT90 in their maximum likelihood analysis, which
will be labelled as Prior 1. The other two are those derived using
Jeffreys' rules described above, i.e. $P_J$ and $P_{J^\prime}$.  These
priors are useful because they provide much more information about $k$
than the LT90 prior.

However, these priors need slight modifications. From equation
(\ref{eq:ltdist}) it can be seen that $k>0$, since negative $k$ would
imply that the probability does not go to zero as $v \rightarrow
\ve$. Therefore for the priors derived from Jeffreys' rules we enforce
that the $P_J = P_{J^\prime} = 0$ for $k<0$. In addition, as can be
seen from equation (\ref{eq:pri_J}), prior $P_J$ diverges as $k
\rightarrow 0$. This is clearly unacceptable because our likelihood
only goes to zero as $k \rightarrow -1$ (equation
\ref{eq:vrdistlt})\footnote{Note that the reason why this likelihood
does not go to zero as $k \rightarrow 0$ is because of the integral
over the unknown tangential velocities performed in equation
(\ref{eq:tan_av})}. In order to avoid this problem we introduce an
additional factor of $k/\sqrt{k+2}$ into prior $P_J$, which means that
$P_J\rightarrow0$ as $k \rightarrow 0$, which is useful for continuity
at $k=0$. Importantly, we still retain the property that
$P_J\rightarrow0$ as $k \rightarrow \infty$.  

All priors are described in Table \ref{sim1.tab}. These different
forms for the {\em a priori} probability distributions apply different
assumptions and weights upon the variables. To further assess these
prior distributions, maximization of equation (\ref{eq:lhood}) using
the parameters $\ve$ and $k$ (refer to Section \ref{sec:boot}) was
performed after applying both of the above priors.

\subsection{Using simulations to assess the choice of prior}
\label{sec:bootsim}

We tested the choice of prior by applying the bootstrap technique to
a simulated dataset. A random sample of stars was drawn from the assumed 
LT90 probability density function (equation \ref{eq:vrdistlt}) for
user-defined values of $\ve=600\kms$, $k=2.0$, and
$\vmin=300\kms$. These values are similar to what we might expect for
our Galaxy. We chose to analyse a random sample size of 35 simulated
stars as this is comparable to the number of stars in our observed
sample (Section \ref{sec:data}). This mock sample was used to test the
bootstrap maximization methods as well as assess chosen priors in
equation (\ref{eq:lhood}) for calculating the escape velocity and
$k$. Note that we apply the LT90 approximation (equation
\ref{eq:vrdistlt}) when calculating the maximum likelihood.

The mock 35 star sample was run through the bootstrap maximization
technique for the three priors given in Table \ref{sim1.tab} applied to
equation (\ref{eq:lhood}). The bootstrap performed maximization on
5000 resamples of the original mock data set.  The bootstrap
distribution of peak likelihood points for $\ve$ is very close to
normal.  Therefore, we used techniques as described in Section
\ref{sec:boot} to compute confidence for $\ve$.
However, the bootstrap distribution for $k$ is clearly not normal. We
determined that the distribution of peak likelihoods follows a
chi-squared distribution with the number of degrees of freedom
equalling the mean of the bootstrap distribution. Therefore
confidence limits for $k$ were derived from a chi-squared distribution
instead of a normal distribution.

Table \ref{sim1.tab} gives the maximization results for each prior
defined in Section \ref{sec:prior_choice}. The third column of the
table represents the maximum likelihood values of $\ve$ and $k$
calculated from the original mock sample before the bootstrap has been
applied. The fourth column give the mean bootstrap values of $\ve$ and
$k$ obtained from the bootstrap distribution. The standard errors (SE)
were computed from the standard deviation of the bootstrap resamples
as explained in Section \ref{sec:boot}. The last column gives the 90
per cent confidence intervals of $\ve$ and $k$. Note, confidence
endpoints for $\ve$ are computed using the bootstrap normal-distribution
approximation, while that for $k$ were computed from a chi-squared
distribution.

Comparing the results from the three different priors, there are two
very noticeable attributes. Firstly, the bootstrap distribution of
$\ve$ from Prior 1 has slightly less bias than that from Priors 2 and
3. This suggests that Priors 2 and 3 are `stronger' priors,
introducing bias into the maximum likelihood analysis.  However, the
bootstrap mean from Prior 2 gives estimates for $\ve$ that are still
identical, within standard errors, to the chosen values for the
simulation.  Prior 3, however, seems to slightly underestimate the
value of $\ve$. The other attribute to notice is that the errors and
confidence regions from Prior 2 are less than that of Prior 1, while
also still containing the desired maximum likelihood values for $\ve$
and $k$.  Therefore it would be useful to use the confidence regions
from Prior 2 when concluding on the confidence of our actual data
sample, especially since it contains more information about
$k$. Notice that Prior 3 has similar confidence regions to Prior 2,
but since the bootstrap mean estimate for $\ve$ is beyond 1 sigma of
the assumed value we believe that Prior 3 may not be as useful.
Therefore we will compute our analysis using Priors 1 and 2 for comparison.

To further assess the reliability of the bootstrap method, we repeated
our analysis on a large number of different mock samples.  These mock
samples were chosen to cover a wide range of values for the input
parameters of our distribution ($\ve \in [400,\,700]$, $k \in
[2,\,5]$).  For only 4 per cent of these samples did the bootstrap
method produce maximum likelihood values beyond $2\sigma$ of the
actual values defined. Otherwise, the technique extracted values of
$\ve$ and $k$ that were consistent within standard errors to their
assumed values in the simulation for all {\em a priori} probability
definitions. Furthermore, the technique was applied using several
different initial guesses for $\ve$ and $k$ during maximization and
the conclusions from all priors remained unaffected.

\begin{table*}
\begin{tabular}{cllccc}
\hline Label & Prior Form & Initial MLE & Bootstrap & SE & 90\%-Conf. \\ 
& & Values & Mean Values & & \\
\hline
1 & $P_{LT}(\ve,\,k) \propto 1/\ve$ & $\ve=631.1\kms$ & 625.6 &
132.8 & (502.7,832.7) \\
& & $k=2.5$ & 2.7 & 2.9 & (0.3,7.3) \\
& & & & & \\
2 & $P_J(\ve,\,k) \propto \frac{\sqrt{k}}{(\ve-\vmin)\,(k+2)}$
 & $\ve=604.6\kms$ & 588.0 & 74.1 & (516.3,730.6) \\
& & $k=2.1$ & 2.0 & 1.5 & (0.1,6.0) \\
& & & & & \\
3 & $P_{J^\prime}(\ve,\,k) \propto \frac{1}{(\ve-\vmin)\,(k+1)}$
 & $\ve=563.5\kms$ & 544.0 & 54.2 & (495.9,634.6) \\
& & $k=1.2$ & 1.0 & 1.1 & (0.0,3.8) \\
& & & & & \\
\hline
\end{tabular}
\caption{Data from the optimization of 35 stars with simulation
 parameters: $\ve=600\kms$, $k=2.0$, $\vmin=300\kms$ after applying
 our derived priors to equation (\ref{eq:lhood}). The third column
 gives the maximum likelihood values of $\ve$ and $k$ for the original
 (non-bootstrapped) sample, while the fourth column gives the mean of
 the bootstrap distribution of maximum likelihood values.  The SE
 column gives the standard errors from the standard deviation of the
 bootstrap distribution. The last column gives the 90 per cent
 confidence intervals for $\ve$ and $k$.  The $\ve$ confidence
 intervals are estimated using the bootstrap normal-distribution approximation,
 while the $k$ intervals are calculated from a chi-squared
 distribution with the number of degrees of freedom equal to the
 bootstrap mean value.  Prior 1 is the same as that used by LT90,
 while Priors 2 and 3 are derived from Jeffreys' rules. All priors are
 0 for $k\le0$. Note that the non-bootstrap analysis employs Prior 1.
 }
\label{sim1.tab}
\end{table*}

\section{The archival high-velocity stars}
\label{archival}

Here we list the archival stars that were used to augment the RAVE
dataset. The stars given in Table \ref{tab:beers} are from the Beers
et al. (2000) catalogue of metal poor stars.

\begin{table*}
\begin{tabular}{lccccccc}
\hline Name & $\vr$ & $\delta\vr$ & RA (J2000) & Dec (J2000) & $\ell$
& b & $V$\\
& ($\kms$) & ($\kms$) & ($^\circ$) &
($^\circ$) & ($^\circ$) & ($^\circ$) & (mag)\\ \hline
BPS CS 30339-0040 & 337.6 & 10  & 5.10900   & -36.50553 & 336.03 & -78.55 & 13.00\\
BPS CS 22166-0024 & -341.2 & 10 & 15.97933  & -12.69761 & 134.99 & -75.28 & 13.86\\
BPS CS 22189-0007 & -307.8 & 10 & 39.39217  & -12.93944 & 188.44 & -61.42 & 13.21\\
BPS CS 22173-0015 & -320.5 & 10 & 61.44904  & -17.35106 & 211.13 & -44.26 & 13.22\\
BPS CS 22177-0009 & -307.2 & 10 & 61.91921  & -25.04522 & 221.74 & -46.17 & 14.27\\
BPS CS 22871-0070 & -311.1 & 10 & 220.31808 & -18.61142 & 336.14 & 37.08 & 14.76\\
BD+01 3070        & -306.2 & 3  & 230.66700 &   1.26469 & 3.87 & 45.48 & 10.06\\
BPS CS 30312-0006 & -342.7 & 10 & 233.24071 &  -1.88931 & 2.76 & 41.49 & 13.65\\
HD 178443         & 336.4 & 10  & 287.65358 & -43.27658 & 354.18 & -21.52 & 9.99\\
BPS CS 22964-0074 & -340.1 & 10 & 297.37250 & -39.71078 & 0.09 & -27.58 & 14.46\\
BPS CS 22943-0087 & 434.7 & 10  & 304.82696 & -46.45764 & 353.38 & -34.05 & 14.24\\
V$\star$ V1645 Sgr& -326.3 & 14 & 305.18525 & -41.11817 & 359.80 & -33.68 & 11.96\\
V$\star$ AO Peg   & 310.8 & 35  & 321.77038 &  18.63386 & 69.94 & -22.58 & 12.83\\
BPS CS 22948-0093 & 396.3 & 10  & 327.63137 & -41.13033 & 0.21 & -50.54 & 15.18\\
BPS CS 22951-0055 & 341.9 & 10  & 328.68683 & -46.52406 & 351.66 & -50.36 & 14.78\\
HD 214161         & -375.3 & 10 & 339.28350 & -40.51067 & 358.40 & -59.31 & 9.10\\
BPS CS 22949-0026 & 313.2 & 10  & 350.75979 &  -5.21428 & 75.14 & -59.61 & 15.23\\
\hline
\end{tabular}
\caption{The archival high velocity stars, taken from Beers et al. (2000). 
The first column shows the Galactocentric radial velocity.
Note that three stars have been excluded from this sample as their
estimated distances are greater than 5 kpc; all stars listed in this
table have estimated distances of less than 2.5 kpc (see Section
\ref{sec:include}).}
\label{tab:beers}
\end{table*}

\end{document}